\documentclass[twocolumn,showpacs,superscriptaddress,preprintnumbers,amsmath,amssymb,prb]{revtex4-1}

\usepackage{graphicx}
\usepackage{dcolumn}
\usepackage{bm}
\usepackage{color}
\usepackage[caption=false]{subfig}
\usepackage{hyperref}
\usepackage{tabularx} 
\usepackage[section]{placeins} 

\begin{document}



\title{Extended spin model in atomistic simulations of alloys \\
 }

\author{Fan Pan}
\email{fanpan@kth.se}
\affiliation{Department of Applied Physics, School of Engineering Sciences, KTH Royal Institute of Technology, Electrum 229, SE-16440 Kista, Sweden}
\affiliation{SeRC (Swedish e-Science Research Center), KTH Royal Institute of Technology, SE-10044 Stockholm, Sweden}
\author{Jonathan Chico}
\affiliation{Department of Physics and Astronomy, Uppsala University, Box 516, SE-751 20 Uppsala, Sweden}
\affiliation{Peter Gr\"unberg Institut (PGI-1)/Institute for Advanced Simulation (IAS-1), Forschungszentrum J\"ulich and JARA, D-52425 J\"ulich, Germany}
\author{Anna Delin}
\affiliation{Department of Applied Physics, School of Engineering Sciences, KTH Royal Institute of Technology, Electrum 229, SE-16440 Kista, Sweden}
 \affiliation{SeRC (Swedish e-Science Research Center), KTH Royal Institute of Technology, SE-10044 Stockholm, Sweden}
\affiliation{Department of Physics and Astronomy, Uppsala University, Box 516, SE-751 20 Uppsala, Sweden}
\author{Anders Bergman}
\affiliation{Department of Physics and Astronomy, Uppsala University, Box 516, SE-751 20 Uppsala, Sweden}
\affiliation{Maison de la Simulation, USR 3441, CEA-CNRS-INRIA-Universit\'{e} Paris-Sud-Universit\'{e} de Versailles, F-91191 Gif-sur-Yvette, France}
\affiliation{INAC-MEM, CEA, F-38000 Grenoble, France}
\author{Lars Bergqvist}
\affiliation{Department of Applied Physics, School of Engineering Sciences, KTH Royal Institute of Technology, Electrum 229, SE-16440 Kista, Sweden}
 \affiliation{SeRC (Swedish e-Science Research Center), KTH Royal Institute of Technology, SE-10044 Stockholm, Sweden}

\date{\today}

\begin{abstract}
An extended atomistic spin model allowing for studies of the finite temperature magnetic properties of alloys is proposed. 
The model is obtained by extending the Heisenberg Hamiltonian via a parameterization from a first principles basis, interpolating from both the low temperature ferromagnetic and the high temperature paramagnetic reference states. This allows us to treat magnetic systems with varying degree of itinerant character within the model. Satisfactory agreement  with both previous theoretical studies and experiments are obtained in terms of Curie temperatures and paramagnetic properties. The proposed model is not restricted to elements but is also applied to binary alloys, such as the technologically important material Permalloy, where significant differences in the finite magnetic properties of Fe and Ni magnetic moments are found. The proposed model strives to find the right compromise between accuracy and computational feasibility for accurate modelling, even for complex magnetic alloys and compounds.    

\end{abstract}

\maketitle

\section{\label{sec:intro}Introduction}


First principles calculations and atomistic modelling play an important role for understanding the properties of magnetic compounds and alloys. In particular, the finite temperature properties determine the usefulness of the material for technological applications. However, a complete theory for finite temperature magnetic properties remains a formidable challenge in condensed matter physics despite the vast advances in the recent decades. 
Such a theory need to include not only transversal but also longitudinal spin fluctuations (LSF) of the magnetic moments in responsible for the moment formation in the material. 
Magnetic moments can roughly be classified in two different types, either being well localized, or exhibiting a more itinerant character. Previous studies have typically described localized moment systems by means of the Heisenberg model\cite{Heisenberg1928} and itinerant magnets by band theories\cite{Bloch1929, Stoner1936, Slater1936, Slater1936b, Stoner1938}, respectively. However, several efforts\cite{Moriya1979, Uhl1996, Rosengaard1997, Ruban2007}  have been directed towards a unification of the existing approaches with the goal of being able to describe the two different scenarios within a single spin model.
 
The first step was taken by Moriya\cite{Moriya1979} describing the magnetic phase transition based on low-energy excitations. In the frame work of Ginzburg-Landau theory, the Hamiltonian is expanded in the even powers of the local moment magnetization $\mathbf{M}(\mathbf{r})$ where the expansion coefficients are obtained by calculating the constrained total energy $E(M)$\cite{Mohn1990}. This theoretical effort was followed by Uhl and K\"ubler\cite{Uhl1996}, who formulated a spin fluctuation theory in which the exchange parameters of the Hamiltonian were calculated from first-principles of spin-spiral states. 
Inspired by this work, Rosengaard and Johansson\cite{Rosengaard1997} proposed a Hamiltonian where a LSF term was introduced in addition to the Heisenberg term and subsequently calculated thermodynamic properties by means of Monte Carlo simulations. Ruban \textit{et al.}\cite{Ruban2007}, developed an extended scheme where a more advanced treatment of the LSF was introduced. In this latter approach, the disordered local moment was used as the reference state of mapping the total energies and thus the model can be expected to give a more accurate description of the paramagnetic state compared to the FM ground state.

A large number of works have later been focused on the effect of longitudinal fluctuations in the description of the half metallic properties of the Heusler alloys\cite{Lezaic2006, Lezaic2013, Sandratskii2007, Sandratskii2008}. Common to these works is the proposal of a model Hamiltonian that considers a separate treatment of the strong-moment components and the induced moment, such that it can effectively take into account the fluctuations on the induced moments by renormalization of the exchange interactions. A similar technique has been applied to systems with multiple sublattices in other studies\cite{Polesya2010, Polesya2016}. A further extension includes calculation of the ordering temperature not only to the ferromagnetic-nonmagnetic transition but also for the N\'{e}el temperature of antiferromagnetic systems\cite{Polesya2016, Khmelevskyi2016}. Parallel to the studies of equilibrium magnetic properties from Monte Carlo methods, there has recently been attempts to include longitudinal fluctuations also in dynamical simulations in the framework of Langevin spin dynamics\cite{Ma2012a, Ma2014}. However, so far it has only been applied for simple elemental metals as Fe and Gd.

Experimentally, the degree of itinerancy is most frequently studied in the paramagnetic susceptibility, either by comparing the effective moment extracted from the Curie-Weiss constant with the true local moment, or the deviation of the Curie-Weiss law. Recently, a great deal of progress has been achieved to study the Stoner excitations through the transversal dynamical susceptibility from first principles within the framework of linear response time dependent density functional theory \cite{Buczek2011,Lounis2015}, or from tight binding theory \cite{Costa2004} . However, due to the complexity of such calculations one faces with difficulties applying the theory to complex magnetic compounds and alloys and to include temperature effects. An issue that often occurs when mapping total energies from first principles to a spin model is that the calculated exchange interactions sometimes render distinctly different results depending on the starting reference state of the mapping. Hence, it would be very attractive to construct an extended spin model where the reference state dependence is eradicated and at the same time include effects of LSF. 

In this paper, we propose a general extension of the Heisenberg model that indeed removes the dependence on the reference state and includes not only the low energy transversal excitations but also the longitudinal spin fluctuations. The scheme combines first principles and Monte Carlo (MC) simulations and provides a more rigorous description of the finite temperature magnetic properties compared to models that includes only transversal fluctuations. In particular, it is expected that a wider range of materials can be investigated within our model, including materials with induced moments.

The paper is organized as follows. Our extended spin model and the procedure of calculating the parameters from first principles are presented in Sec.~\ref{subsec:formalismOfHm}. The modified algorithm in the Monte Carlo for updating the moments in each step is clarified in Sec.~\ref{sec:alsf}. Some practical issues including the approximations made from the mapping of the system on the extended spin model are discussed in Sec.~\ref{subsec:details} and \ref{subsec:totalE}. In the numerical results, Sec.~\ref{subsec:Tc} we compare the critical temperature of each of the considered systems and study the influence of imposing the LSF term. The energy and moment distributions as functions of temperature are shown in Sec.~\ref{subsec:Edistribution} and Sec.~\ref{subsec:Alm}. Paramagnetic properties, specifically the magnetic short range order (MSRO) are discussed and compared with other works in Sec.~\ref{subsec:MSRO}. Sec.~\ref{sec:remarks} contains technical remarks and comparisons to existing methods found in the literature and we conclude and summarize our work in Sec.~\ref{sec:summary}.\\

\section{\label{sec:methods}Formalism}
\subsection{\label{subsec:formalismOfHm}
Formulation of the problem} 

\begin{figure}[htb]
\includegraphics[width=7cm]{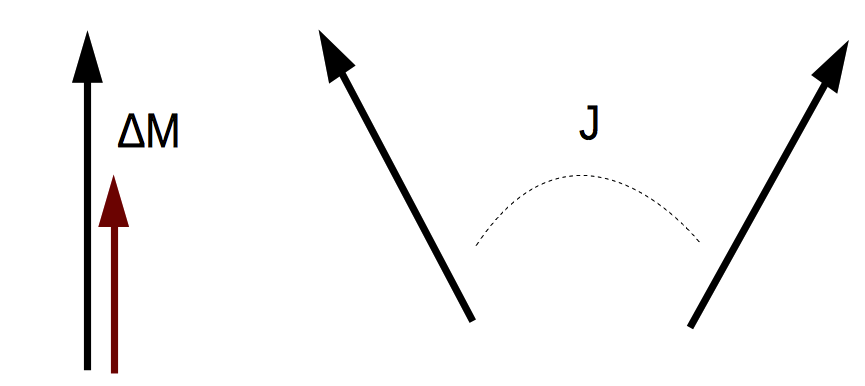}
\caption{Schematic illustration of the two types of excitations possible for a magnetic material. The left panel illustrates longitudinal spin fluctuation (change of moment size $\Delta M$) of each magnetic moment and the right panel illustrates transversal spin fluctuations of two moments interacting with strength J.}
\label{fig:Hextend2}
\end{figure}

A magnetic material at finite temperatures displays two different types of excitations, longitudinal and transversal fluctuations, as illustrated in Fig.~\ref{fig:Hextend2}. 
The starting point for the simulations is the construction of a model Hamiltonian that includes both the longitudinal and transversal spin fluctuations of the magnetic moments $\mathbf{m}$, such that the total Hamiltonian can be expressed as

\begin{equation}
    H_{tot}=H_0+H_{LSF}(\{|\mathbf{m}_i|\})+H_2(\mathbf{m}_i,\mathbf{m}_j),
    \label{eq:Hextend}
\end{equation}
where $\mathbf{m}_i$ denotes magnetic moment at site $i$, $H_0$ is the nonmagnetic reference energy, $H_{LSF}$ and $H_2$ the energies (Hamiltonian) of longitudinal and transversal spin fluctuations, respectively. The nonmagnetic reference energy can be disregarded from the problem since it only sets a reference energy which does not change during the simulations. The proposed form of the Hamiltonian, Eq.~(\ref{eq:Hextend}) takes the form

\begin{equation}
\begin{split}
H_{tot}= & \sum_i  J_1(\{|\mathbf{m}_i|\}) \\ & -  \sum_{ij}(1+k) J_{ij}(|\mathbf{m}_i|,|\mathbf{m}_j|)\mathbf{m}_i \cdot \mathbf{m}_j.
\end{split}
\label{eq:Htot}
\end{equation}
In Eq.~(\ref{eq:Htot}), the first term describes the total energy of the homogeneous paramagnetic state which in the model corresponds to the LSF energy that in turn depends on the parameter $J_1$ which depends on the magnitudes of the local moments of each magnetic atom type of the system. The second term is the standard Heisenberg Hamiltonian of transversal fluctuations with the important distinction that the exchange parameters $J_{ij}$ now are explicitly dependent on the size of the magnetic moments and ``rescaled" with the parameter $k$, as will be explained below. All the parameters entering the Hamiltonian are in our model parameterized from first principles calculations from which the total energy is mapped to the Hamiltonian. When dealing only with transversal fluctuations, a common weakness is the reference state dependence of the calculated exchange interactions\cite{Heine1990}. This weakness does not come as a surprise since the Heisenberg model assumes rigid moments where the size of the magnetic moment does not change upon rotation, \textit{i.e.} strictly speaking only valid for localized spins. Unfortunately, most of the real materials possess some degree of itinerancy where the size of the moment changes depending on internal orientation rendering Heisenberg model less applicable. For instance, one severe and common example is Ni where the magnetic moment vanishes for tilting angle between two nearest neighbour moments of around 60 degrees or more \cite{Rosengaard1997,Halilov1998}.

\begin{figure}[htb]
\includegraphics[width=8cm]{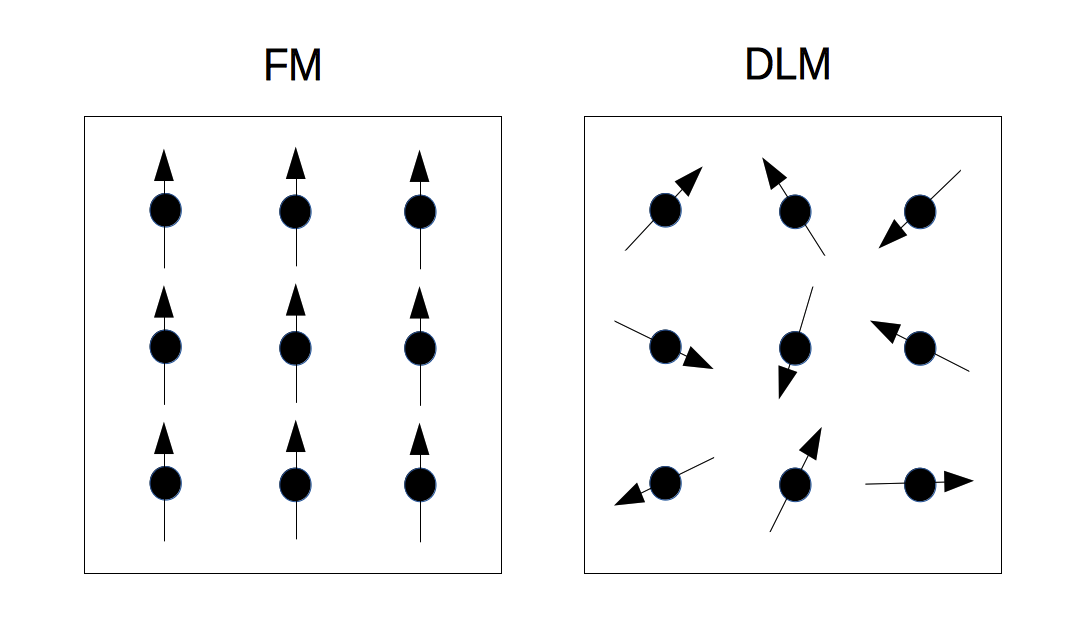}
\caption{Illustration of the reference states used in the mapping of the extended spin Hamiltonian from DFT. FM denotes the low temperature ferromagnetic state and DLM the high temperature disordered local moment state.}
\label{fig:refstates}
\end{figure}

Another large important class of materials for which a pure Heisenberg description fails to describe the properties correctly are systems which exhibit induced moments, \textit{i.e.} moments that only appear in the presence of other localized moments creating a local internal magnetic field that polarizes the atoms. Even for systems where the Heisenberg description works reasonablely well, such as Fe, the calculated exchange interactions do depend on the reference state \cite{Heine1990,Szilva2013} and hence the calculated critical temperatures could differ to some extent. Szilva {\it et. al.}\cite{Szilva2013} proposed a formalism that allows the calculation of the exchange interaction from a general non-collinear magnetic state that better represents the configurations at finite temperatures by adding a biquadratic term. Here, in this study we take an alternative approach to soften the reference state dependence and at the same time include longitudinal spin fluctuations. This is achieved by using the extended spin Hamiltonian, Eq.~(\ref{eq:Htot}) and employing two reference states in the mapping, as illustrated in Fig.~\ref{fig:refstates}. The most natural reference states are the ferromagnetic (FM) low temperature state and the paramagnetic high temperature state. The latter is referred as the disordered local moment (DLM) state. Both states are naturally treated in a straight forward manner in the present computational framework.  

The proposed Hamiltonian Eq.~(\ref{eq:Hextend}), is designed in such a way that it is exact for both the FM and DLM states, in the sense the total energies from DFT are recovered, and it interpolates between the two limits at intermediate temperatures. This is most easily realized in the DLM limit where the Heisenberg term is equal to zero such that the only surviving term is the sum of $J_1$, consistent with the model proposed by Shallcross {\it et. al} \cite{Shallcross2005} and Ruban {\it et. al} \cite{Ruban2007}. The parameter $k$ can then be viewed as a rescaling parameter that tunes the interpolation between the DLM and FM limits. However, it is worth stressing that the model does not contain any free parameters. Calculating exchange parameters starting from either FM or DLM state will indeed result in different sets of $J_{ij}$. However, since the overall Hamiltonian is fixed, it will result in different values of the rescaling factor $k$ such that the ``rescaled" exchange parameters as well as the exchange energy are similar, independently from the starting reference state. 
The most accurate LSF term in the Hamiltonian describes the energy cost or gain of forming a local magnetic moment with size $|\mathbf{m}|$ embedded in a host with the average moment $\langle m \rangle$. This choice was employed by Shallcross {\it et. al} \cite{Shallcross2005} and Ruban {\it et. al} \cite{Ruban2007} but has the main disadvantage that it becomes prohibitively complicated for systems with several magnetic components, such as binary and ternary alloys. An approximation to simplify the model is to assume that the embedded moment has the same size as the host, in such a way that the number of indices is reduced to the number of components without the additional index of the average moment. For one atom per cell, this becomes the well known Landau expansion, in which the energy is expanded in even powers of the size of the magnetic moment $|\mathbf{m}|=m$, i.e. $E(m) \approx a_1m^2+a_2m^4 + ...$, where $a_i$ are expansion coefficients. By employing constrained density functional theory using the Fixed Spin Moment (FSM) technique, $E(m)$ is easily calculated and analytically fitted to the Landau expansion expression. Even with this simplified LSF term in the Monte Carlo simulations, a number of studies\cite{Rosengaard1997, Lezaic2006, Ma2012a} has been demonstrated with rather satisfactory results for systems with one atom per cell, such as Fe. 

Alternative methods to treat longitudinal fluctuations include the spin cluster expansion (SCE) technique \cite{Drautz2004}, in which the magnetic energy from DFT is rigorously expanded in series of multispin interactions. For the case of a weak induced moment in presence of a strong moment, specialized schemes have been designed by Lezaic {\it et. al} \cite{Lezaic2013} and Polesya {\it et. al} \cite{Polesya2010}.    

However, as far as we are aware, there has not been any systematic attempts to expand the formalism to the general case with several magnetic components, which is the main motivation for the present work. To be more specific, we here expand the formalism to binary alloy systems. This covers many of the technologically important alloys but also the essential class of materials consisting of strong and weak magnetic moments, where the weak moment is induced by the strong moment from an internal magnetic field. The extension to several magnetic components is straight-forward, although the computational effort drastically increases. Nevertheless, by employing modern computational techniques as discussed below it is feasible without too much computational effort. The crucial step in the parametrization of the full LSF energy surface is to constrain the size of the magnetic moment on one component at the time. For simplicity, we restrict the discussion to binary alloys with magnetic components $\mathbf{m}^\alpha$ and $\mathbf{m}^\beta$, respectively. This could, as an example, represent the Fe and Co magnetic moments in a FeCo alloy. To obtain reasonable numerical accuracy, around 20 different moment sizes of each component are required, such that the total number of configurations is $20^2=400$ to fully parametrize the LSF energy surface. However, each configuration is independent of each other, so clever scripting and making use of the harvesting power of modern parallel computers makes the parametrization efficient despite the vast number of total configurations required. The moment sizes are arranged in a $n \times m$ Cartesian grid with indices $(m_i^\alpha,m_j^\beta)$, where $\alpha=\{1,..n\}$ and $\beta=\{1,...,m\}$. n and m are the number of moment sizes performed for $m_i$ and $m_j$, respectively. Indices $i$ and $j$ denote the site indices. Electronic structure calculations based on density functional theory for each grid point yield the input parameters $J_1,k$ and $J_{ij}$ in the Hamiltonian, Eq.~(\ref{eq:Htot}), for that particular grid point.

\subsection{Atomistic simulations including longitudinal moment fluctuations}\label{sec:alsf}

Following the full Hamiltonian parametrization, one can use a statistical treatment, for instance Monte Carlo (MC) simulations to work out the equilibrium properties or atomistic spin dynamics (ASD) for the dynamical properties. For systems with one magnetic atom per cell, MC simulations have been performed in a number of previous
studies\cite{Rosengaard1997,Lezaic2006,Ruban2007,Lezaic2013} while ASD simulations are much less explored\cite{Ma2012a,Ma2014}. In this study, we restrict ourselves for the moment to MC simulations while ASD simulations are left for a future study. Using the Metropolis algorithm, each trial move of every selected atomic moment not only consists of the normal rotation but also a random change of the magnitude of the moment. Since the Hamiltonian is only defined at the grid points, we have employed bilinear interpolation of the intermediate points in the simulations. As mentioned above, we are using a slightly simplified LSF Hamiltonian and it has some practical consequences. When the magnitude of a trial moment is changed in the Metropolis update, \textit{i.e.} $m_i^{\alpha'}=m_i^\alpha+\delta m$, we make the approximation that we use the average magnitude $\langle m^\beta\rangle$ of the other component in the binary alloys around the site $i$, 
such that the parameters $J_1,k$ and $J_{ij}$ yield the configurations corresponding to $(m_i^\alpha,\langle m^\beta \rangle)$ and after the change $(m_i^{\alpha'},\langle m^\beta \rangle)$, respectively.
This approximation does not cause any significant issues for the homogeneous systems since the magnitude of magnetic moments in average does not change much within a sphere from the trial atom point of view. Comparing this approximation against doing interpolation of $J_{ij}$ between each pair $m_i$ and $m_j$, only minor differences were found but it gains a magnitude faster in the computational efforts.
The $J_1$ term may be a bit more sensitive since it is assumed that moments belonging to the same component $\alpha$ change the moment size instantaneously with the trial moment. Comparison of different strategies are discussed in more detail in Sec.~\ref{sec:remarks}.

\subsection{\label{subsec:details} Details of the calculations} 

Electronic structure calculations based on density functional theory (DFT) were employed for extracting all material dependent parameters in the Hamiltoninan. We used an implementation of multiple-scattering theory based on the Korringa-Kohn-Rostoker (KKR) formalism \cite{Ruban2002}. We employed both the local spin density approximation (LDA) and the generalized gradient approximation (GGA) using the PBE parametrization\cite{Perdew1996} as the exchange-correlation potential. Another method normally referred to as Fixed Spin Moment (FSM) \cite{Dederichs1984,Schwarz1984} where the magnitude of the magnetic moment was varied by employing a small constraining field with strength adjusted in such a way that it reaches the prescribed target. Around 20 different moment sizes were used for each magnetic component. The chemical and magnetic disorder were treated using the coherent potential approximation (CPA). A binary alloy with components A and B can be written as $A_xB_{1-x}$, where $x$ is the concentration of A and this configuration is used for the ferromagnetic state (FM). The disordered local moment (DLM) configuration is obtained by forming a four-component alloy, $A^\uparrow_{x/2}A^\downarrow_{x/2}B^\uparrow_{(1-x)/2}B^\downarrow_{(1-x)/2}$, with $\uparrow$ ($\downarrow$) denoting moments pointing along (opposite) to the quantization axis. The exchange interactions $J_{ij}$ were calculated using the LKAG formalism \cite{LKAG1984,LKAG1987}. The experimental lattice constants were used for all systems neglecting any thermal expansion. Moreover, the electronic entropy effects were neglected to avoid additional temperature dependence of the parameters. 

The atomistic Monte Carlo simulations making use of the Metropolis algorithm were performed and implemented in the Uppsala Atomistic Spin Dynamics (UppASD) software package \cite{Skubic2008,ASDbook2017}. 
In each trial move in the Metropolis update, in order to speed up the calculations, one of the three modes was randomly selected between a pure transversal, a pure longitudinal or a combination of the two. 
The transversal updates were in turn chosen randomly to either an uniform rotation, a Gaussian distribution rotation or a spin flip for further speed up. Critical temperatures were extracted using the fourth-order size-dependent Binder cumulant\cite{Binder1981} and since we are dealing with random alloys, for each system and temperature, an ensemble averaging over around 10 different disorder configurations were performed. Calculation of spin stiffness followed the procedure outlined in Ref.[\onlinecite{Pajda2001}] generalized to random alloys using the same methode as in Ref.[\onlinecite{Durrenfeld2015}] averaged over 1000 disorder configurations.       

Evaluation of thermodynamic properties for a model including longitudinal fluctuations involves technically intricate details. The longitudinal part of the partition function over which a functional integration is performed can be written as\cite{Murata1972,Sandratskii2008,Wysocki2008,Dietermann2012,Ruban2013,Sergii2016}

\begin{equation}
    Z=\int g(m) \mathrm{d}m \ \mathrm{exp}\left(-\frac{H_{tot}}{k_BT}\right),
\end{equation}
where $g(m)$ is the phase space measure (PSM)\cite{Wysocki2008}. In this study we use $g(m)=1$, that is the so-called uniform PSM or Murata-Doniach metric that corresponds to the decoupled treatment of longitudinal and transversal spin fluctuations. However, as was pointed out in Ref.[\onlinecite{Wysocki2008}], magnetic moments are not canonical variables and therefore the PSM is not known. Including PSM with Jacobian factor, $g(m)=m^2$, couples the longitudinal and transversal spin fluctuations and it has been argued that this choice could lead to improved results\cite{Sandratskii2008,Ruban2013}. Although not used for the majority of simulations in this study, we did implement this PSM in our MC program and we will briefly discuss some results and differences to the uniform PSM in Sec.~\ref{sec:remarks}.   

\section{Results}
For the results in this section, we consider the elemental materials Fe, Co and Ni, where Fe crystallizes in the bcc structure and Co and Ni in fcc structure. Moreover, we investigate the Permalloy in the fcc structure, which is a binary alloy with composition 81\%Ni and 19\% Fe. Finally, we investigate a binary alloy of Fe and Co in the bcc structure, with Co concentrations ranging from 30\% to 70\%. 

\subsection{\label{subsec:totalE}The total energy landscape from the first-principles calculations}
\begin{figure*}[htb!]
\centering
\includegraphics[clip,width=1.95\columnwidth]{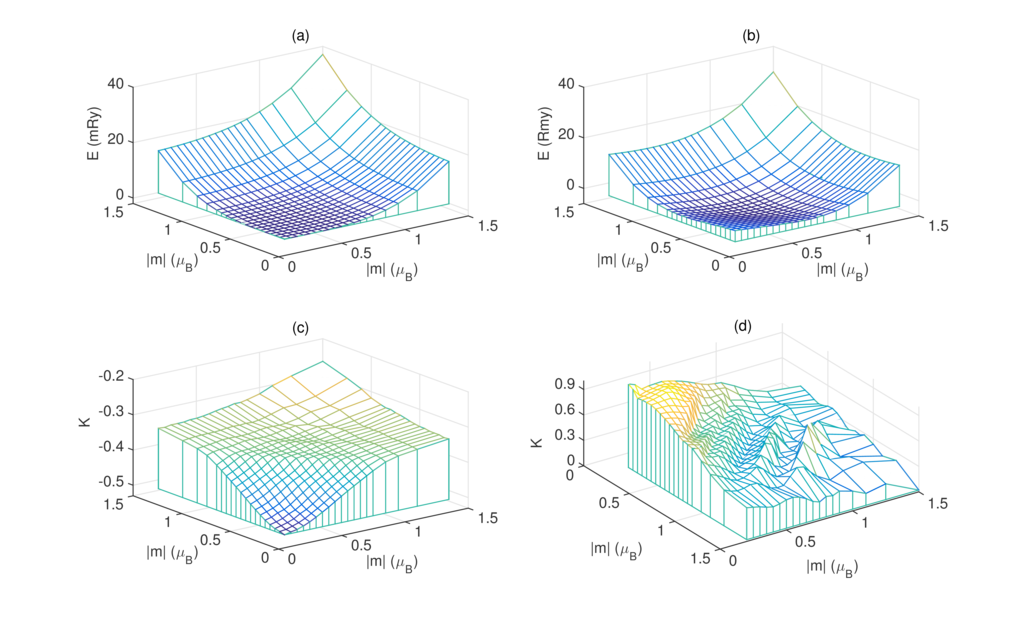}
\caption{The total energy surface of Ni is calculated using the binary-components scheme at each discretized grids of the moments size from two starting points: (a) DLM reference state ($J_1$) and (b) FM reference state. The magnitude of the rescaling factor $k$ is shown in Fig. (c) (DLM) and (d) (FM). Note that for visualization clarity the scale on the z-axis and the viewing angle are different between Fig. (c) and (d).}
\label{fig:FM_DLM_3D_energy_landscape}
\end{figure*}
%
The energy surfaces relevant to the parameters $J_1$ and rescaling factor $k$ are shown in Fig.~\ref{fig:FM_DLM_3D_energy_landscape} for Ni. 
We show that the starting point of using different sets of exchange parameters from DLM and FM state can be eliminated using the rescaled interactions.
Figure~\ref{fig:FM_DLM_3D_energy_landscape}a and \ref{fig:FM_DLM_3D_energy_landscape}b indicate that the total energy of FM state has a shallow minimum of the moment size around 0.7 $\mu B$, while there is no such a minimum in the DLM state, indicating that in the DLM state no magnetic moments can be sustained unless additional fluctuations are present in the system. The total exchange energy is not explicitly plotted here, but from Eq.~(\ref{eq:Htot}) it is apparent that it is the total energy difference between the FM and DLM states. Inspection of individual nearest neighbour exchange interaction corresponding to the total energy minimum for FM, it is found that the DLM reference state gives approximately two times larger $J_{ij}$ than in the FM state. This fact is well reflected on the calculated Curie temperature based on the traditional Heisenberg model.
In Fig.~\ref{fig:FM_DLM_3D_energy_landscape}c and \ref{fig:FM_DLM_3D_energy_landscape}d, the rescaling factor $k$ of the exchange interactions are displayed in the case when the ``bare" exchange interactions are calculated from the DLM and FM states, respectively. A negative (positive) value of $k$ corresponds to reduction (addition) of the exchange such that it forces the exchange energy being the same regardless of chosen initial reference state. Calculating $J_{ij}$ from the FM state gives a larger surface roughness of $k$ compare to calculation from the DLM state due to larger numerical sensitivity, however since the overall total energy surface is smooth, it is not expected to cause any particular numerical difficulties in the simulations. Although we show only Ni as an example of the the energy parameterization, the procedure is similar for the other systems with the key feature that the overall spin model gives correct total energy in both the FM and DLM limits.

\subsection{\label{subsec:Tc}Critical temperatures and spin stiffness}

\begin{table*}
\begin{tabular*}{.85\linewidth}{@{\extracolsep{\fill}}c c c c c| c c c c |c c}
\toprule

T$_c$ (K) &\multicolumn{4}{c|}{DLM} & \multicolumn{4}{c|}{FM} & \multicolumn{2}{c}{Refs.}\\
& \multicolumn{2}{c}{LDA} & \multicolumn{2}{c|}{GGA} & \multicolumn{2}{c}{LDA} & \multicolumn{2}{c|}{GGA} &  &\\
  
System & N & LSF & N & LSF  & N & LSF & N & LSF & Expts. & Theory\\
\hline
Fe & 1132 & 926 & 1275 & 1068 & 1238 & 939 & 1412 & 1083 & 1043$^a$ & 1060$^b$,1065$^c$,900$^d$,1095$^e$\\
Co & 1441 & 807 & 1611 & 979 & 1475 & 791 & 1637 & 902 & 1388$^a$  &1080$^b$,1280$^d$,1012$^e$ \\
Ni & 509 & 432 & 580 & 478 & 486 & 431 & 578 & 459  & 633$^a$ & 510$^b$,615$^c$,412$^e$\\
Py & 645 & 482 & 747 & 572 & 706 & 484 & 761 & 560 & 850$^f$ &650$^g$\\
Fe$_{0.3}$Co$_{0.7}$ & 1564 & 927 & 1686 & 1105 & 1605 & 902 & 1733 & 1080 & - & 1490$^d$ \\
Fe$_{0.5}$Co$_{0.5}$ & 1650 & 1005 & 1862 & 1215 & 1634 & 982 & 1885 & 1165 & 1253-1370$^h$ & 1600$^d$ \\
Fe$_{0.7}$Co$_{0.3}$ & 1689 & 1060 & 1879 & 1267 & 1656 & 1010 & 1847 & 1211 & - & 1490$^d$ \\
\toprule
$^a$ Reference \onlinecite{Wohlfarth1980} \\
$^b$ Reference \onlinecite{Rosengaard1997} \\
$^c$ Reference \onlinecite{Ruban2007} \\
$^d$ Reference \onlinecite{Lezaic2007} \\
$^e$ Reference \onlinecite{Uhl1996} \\
$^f$ Reference \onlinecite{Landolt1991} \\
$^g$ Reference \onlinecite{Kudrnovsky2008} (RPA) \\
$^h$ Reference \onlinecite{Kawahara2003}

\end{tabular*}
\caption{Curie temperatures T$_c$ in K from calculations and comparison to experiments (Expts.) and previous studies (Theory). N (LSF) denotes calculations without (with) longitudinal fluctuations and rescaled exchange parameters. DLM and FM denotes from which reference state the exchange interactions are calculated.}
\label{table:tc}
\end{table*}

In Tab.~\ref{table:tc}, our calculated Curie temperatures, $T_c$ together with previously published experimental and theoretical results are collected. In all of our calculations, the rescaled exchange parameters were used and we compare the two cases when LSF is included or not in the Hamiltonian from Eq.~(\ref{eq:Htot}). A few general trends of the calculated results are immediately noticed. First of all, the inclusion of LSF always lowers the calculated $T_c$ values and underestimates them compared to experiments, with the exception of Fe where excellent agreement is found. Secondly, within the same functional (LDA or GGA), there is only a minor difference in $T_c$ starting from either DLM or FM configuration, which is one of the design goals of the model.  Without LSF and rescaling of the exchange parameters, the well known dependence of chosen reference state is evident. For example, calculated $T_c$ of Ni in GGA is around 400 K (900 K) with ``bare" exchange parameters calculated from FM (DLM) reference, compared to around 580 K using rescaling for both reference states. For the DLM state, the moments were fixed to the same value as in FM, since without constraints or inclusion of LSF term the DLM moment vanishes for Ni. In general, for all the considered systems, without rescaling the exchange interactions from FM configuration underestimate $T_c$ while there is an overestimation using exchange interactions from DLM. Within the same level of approximation and methodology, our calculations corresponds in general rather well with previous published results.

The quantitatively underestimation of $T_c$ with LSF included is however slightly disappointing but could be understood in the present model from relatively simple arguments. If the LSF energy would not manifest a minimum, as in Ni for example, or having a shallow minimum like as in Co, then the system may gain energy of by reducing the moment at finite temperature. However, at the same time, the exchange energy is decreasing upon decreasing moment and fluctuations. On the opposite, increasing moment size gains exchange energy but costs LSF energy, so there is a delicate energy balance between the two contributions. Consequently, with LSF included it does provide the system an alternative path of lowering the energy thus making the system more magnetically ``soft". It could well be that the softening of the moment in the present model is slightly overestimated since the LSF energy is only an approximation to some extent and depends on the average magnetic moments of the local environment. 
Moreover, using the uniform phase space measure as done here causes an oversampling of configurations with smaller moments due to the  decoupling of the transversal and longitudinal fluctuations. 

Fe$_x$Co$_{1-x}$ alloys in bcc structure are known for having a maximum average moment around $x\approx 0.7$ according to the Slater-Pauling relationship, which is found both in experiments \cite{Landolt1991} and from calculations\cite{Abrikosov1996}. These alloys have among the largest Curie temperatures of any discovered transition metal alloy systems, even larger than the individual host elements Fe and Co. This is mostly confirmed by our calculations, with a possible exception of Fe$_{0.3}$Co$_{0.7}$ in GGA that it is very close to the elemental Fe. On a technical note, the results of Fe, Co and Ni in Table~\ref{table:tc} were obtained treating the system as having two components. This makes the evaluation of the exchange and LSF energies slightly more accurate than treating the system as having single component, despite describing the same element. A comparison between the two are discussed in more details in Sec.~\ref{sec:remarks}.

In Tab.~\ref{table:stiffness}, results of our calculated values of the spin stiffness together with experimental and previous published values are compared. Spin stiffness is sensitive to the individual exchange interactions just as the sum of all exchange interactions gives the exchange strength which is relevant to the Curie temperature. The difference being that the spin stiffness contains an additional spatial dependence in the sum. The calculated spin stiffness values are in general in reasonable agreement with both experiments and previous calculations for systems where such data exist. Fe is found to have too large spin stiffness when the exchange parameters are calculated from the FM configuration. This is due to the fact that the individual exchange parameters are not only rather different in DLM or FM configuration, but also rather sensitive to the volume and the basis set in this case. Although the exchange interactions are scaled uniformly to yield the correct total energy, that does not universally translate to improving the spin stiffness. The values of Ni are also overestimated compare to experiments, as found in previous calculations, but the most likely explanation is the use of LDA or GGA that fails to fully describe the electronic structure of Ni being a moderately correlated metal.

In the case of random alloys, Py and Fe$_x$Co$_{1-x}$, the spin stiffnesses were calculated as an average over 1000 different disorder configurations. Perhaps not so surprising, the spin stiffness for Py is rather similar to Ni while values for the Fe$_x$Co$_{1-x}$ alloys are similar to those of elemental Fe and Co. Regarding experiments, inelastic neutron scattering experiments is considered the most straight forward way to measure spin waves and thus the spin stiffness. For the elemental materials Fe, Co and Ni, there are plenty of existing experimental studies and they all rather much agree to each other. However, for random alloys such as Py and Fe-Co, we find only very few studies of direct measurements of $D$. For the measured Fe-Co systems, a thin film geometry was used which could affect the results and that could in turn explain the difference from the calculated bulk value.



\begin{table}[htb]
\begin{tabular*}{.99\linewidth}{@{\extracolsep{\fill}}c c c| c c |c c}
\toprule
D (meV\AA$^2$) &\multicolumn{2}{c|}{DLM} & \multicolumn{2}{c|}{FM} & \multicolumn{2}{c}{Refs.}\\
System & LDA & GGA & LDA & GGA & Expts. & Theory \\

\hline
Fe & 320 & 368 & 466 & 573 & 314$^a$ & 247$^b$,250$^c$\\
Co & 614 & 676 & 675 & 723 & 510$^d$ & 502$^b$,663$^c$\\
Ni & 895 & 924 & 707 & 924 & 550$^e$ & 739$^b$,756$^c$\\
Py & 655 & 693 & 620 & 611 & 390$^f$ & -\\
Fe$_{0.3}$Co$_{0.7}$ & 552 & 600 & 611 & 701 & 476$^g$ & - \\
Fe$_{0.5}$Co$_{0.5}$ & 536 & 588 & 526 & 646 & 800$^g$ & -\\
Fe$_{0.7}$Co$_{0.3}$ & 496 & 536 & 466 & 472 & 470$^g$ & -  \\
\toprule
$^a$ Reference \onlinecite{Stingfellow1968} \\
$^b$ Reference \onlinecite{Rosengaard1997} \\
$^c$ Reference \onlinecite{Pajda2001} \\
$^d$ Reference \onlinecite{Wohlfarth1980} \\
$^e$ Reference \onlinecite{Mitchell1985} \\
$^f$ Reference \onlinecite{Landolt1991} \\
$^g$ Reference \onlinecite{Liu1994} 
\end{tabular*}
\caption{Spin stiffness D in (meV\AA$^2$) from calculations and comparison to experiments (Expts.) and previous studies (Theory). DLM and FM denotes from which reference state the exchange interactions are calculated and LDA and GGA denote different exchange-correlation potentials in calculations.}
\label{table:stiffness}
\end{table}

\subsection{\label{subsec:TandLatft}Finite temperature transversal and longitudinal fluctuations}

\begin{figure}[htp]

\subfloat{\includegraphics[clip,width=0.85\columnwidth]{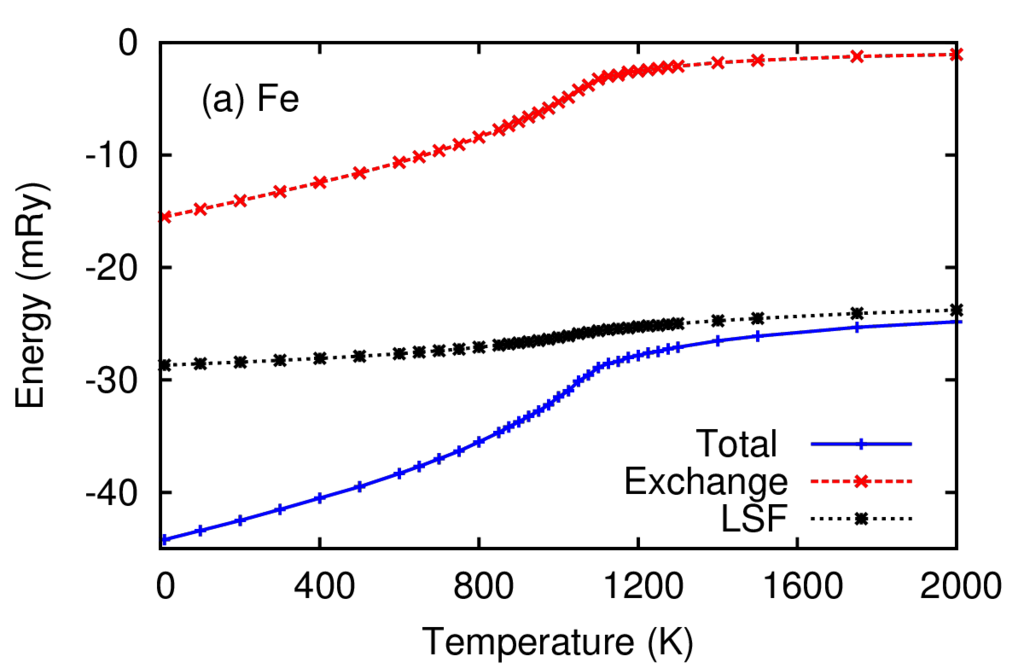}%
}

\subfloat{\includegraphics[clip,width=0.85\columnwidth]{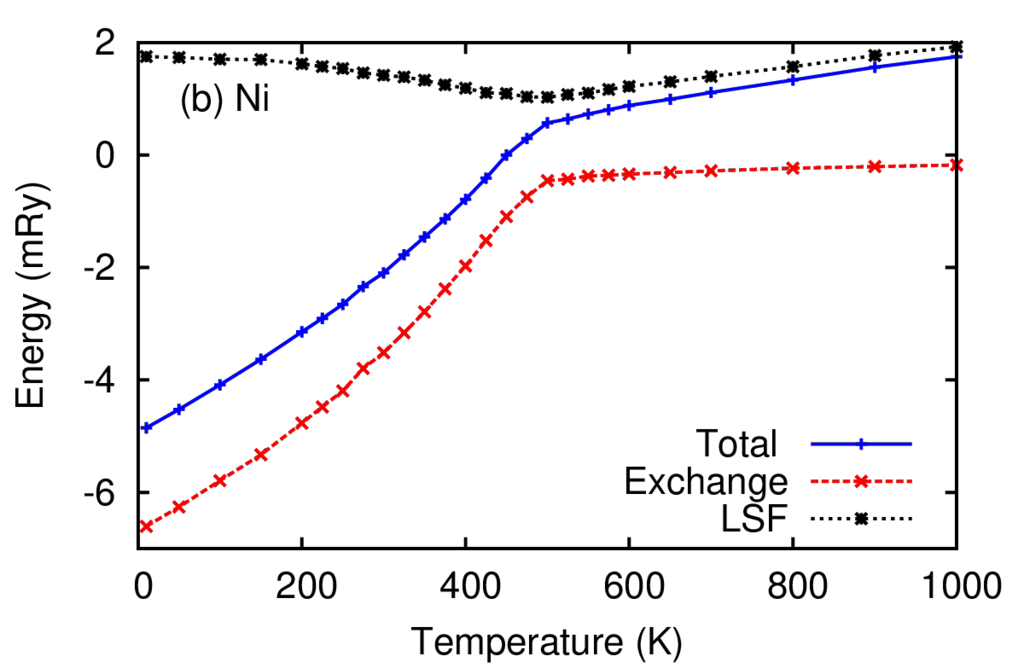}%
}

\subfloat{\includegraphics[clip,width=0.85\columnwidth]{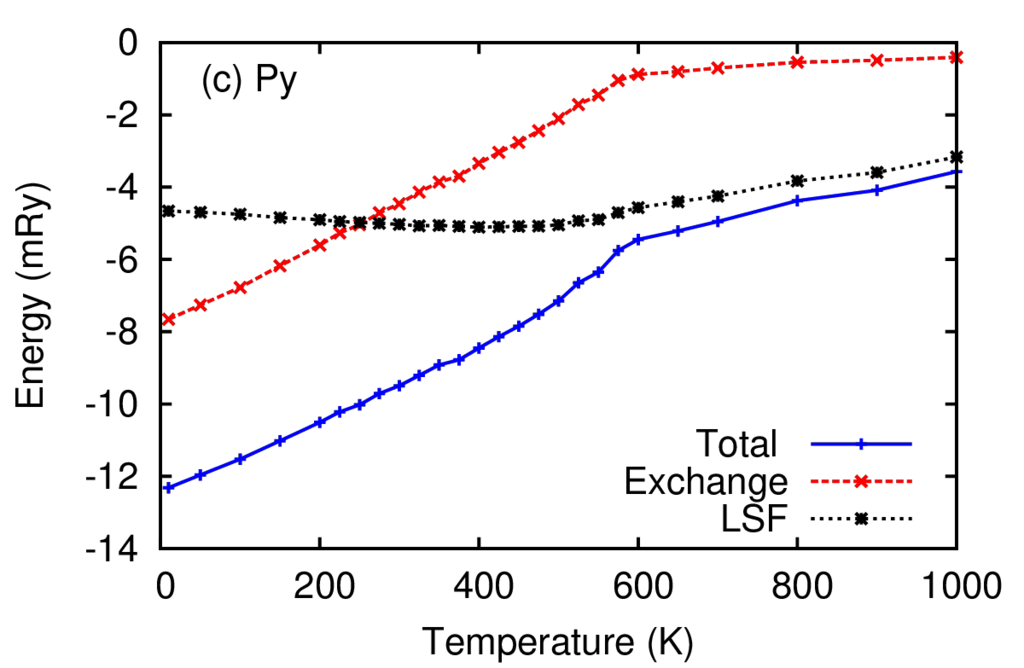}%
}
\caption{Total energy decomposition from Monte Carlo simulations in terms of the total energy, the exchange energy and the LSF energy as a function of temperature for (a) Fe, (b) Ni, and (c) Py.}

\label{EnergyAtFinite_T}
\end{figure}

To further elaborate on the temperature induced magnetic phase transition, we show in Fig.~\ref{EnergyAtFinite_T} the total energy decomposition in terms of the exchange and LSF energies, as a function of the temperature. The exchange term gives the energy gain by aligning moments with each other, while the LSF energy is a measure of the formation (reduction) of a local moment. By resolving the exchange and LSF energies, the stability of the local moment at a certain temperature can be estimated. In the case of Fe, Fig.~\ref{EnergyAtFinite_T}a, the LSF energy is more or less constant, while the exchange energy is monotonous increasing with respect to the temperature and vanishes at high temperature where all moments are randomly distributed. 
The energies confirm that Fe is forming a stable moment in both the FM and DLM states.  
In contrast to Fe, Ni as shown in Fig.~\ref{EnergyAtFinite_T}b reveals a much more precarious energy balance. Upon increasing the temperature to around $T_c$, a part of the exchange energy increase is compensated by decreasing the LSF energy, causing an reduction of the magnetic moment.
For Py, shown in Fig.~\ref{EnergyAtFinite_T}c,  the interplay of the LSF and exchange energies behaves similar as to elemental Ni where part of the loss of the exchange energy is compensated by the LSF energy. As a result, the moments of Py in average have a tendency to be reduced at finite temperatures.

\subsection{\label{subsec:Edistribution}Energy distribution}

\begin{figure}[htp!]

\centering \includegraphics[clip,width=1.00\columnwidth]{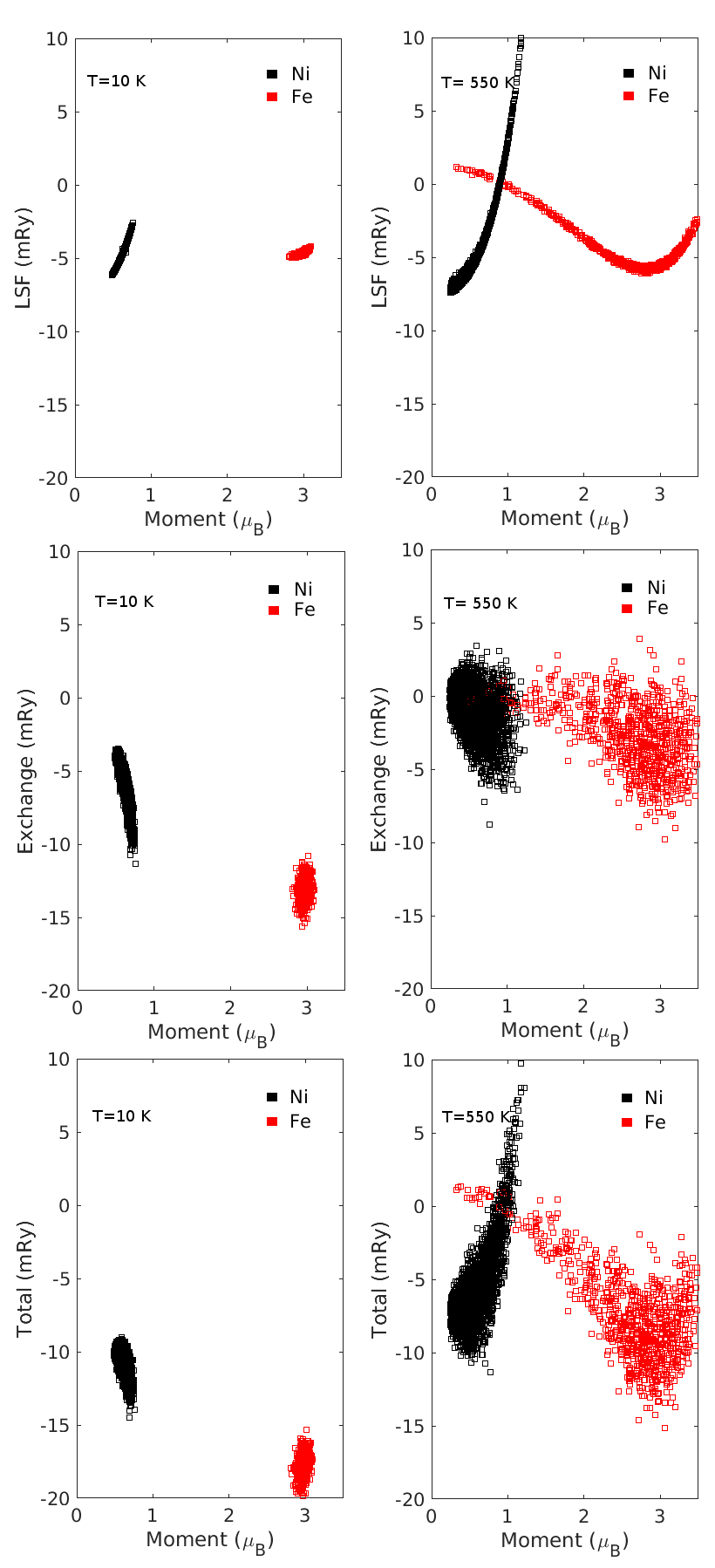}%

\caption{Site resolved energy decomposition (LSF, exchange and total energy, top to bottom) of Py from a snapshot configuration using Monte Carlo simulations at $T=10\ K$ (left) and $T=550\ K$ (right)}
\label{fig:Pyedisp}
\end{figure}

In the previous section we discussed the average decomposed energies as function of temperature. Here, we study in more detail the distribution of energies during the simulations.  

In Fig.~\ref{fig:Pyedisp}, the composition of site resolved energies of Py from a snapshot during the Monte Carlo simulations is displayed for two temperatures, at low temperature ($10\ K$) and around Curie temperature ($550\ K$, see Tab.~\ref{table:tc}). If the energies are averaged both over all the atoms and in time, the results in Fig.~\ref{EnergyAtFinite_T} are recovered. The spread of the magnitude of moments are discussed in more details in Sec.~\ref{subsec:Alm}.  

At low temperatures, the thermal fluctuations are very small and consequently the spread of energies are small. The total energy is fluctuating around the minimum of the energy surface and the largest difference between Ni and Fe moments is the larger exchange energy associated with Fe. At elevated temperature close to the Curie point the situation is rather different. First of all, due to the large thermal fluctuations the spread of both energies and moment magnitudes are significant. Also the directions of moments have large spread and since the figure shows a particular snapshot, there are certain moments that are anti-aligned to each other causing positive site exchange energies. All these fluctuations makes the system different from pure DLM configurations where the moment distribution are random but with fixed magnitude of moments. The high temperature snapshot configurations could be viewed as many DLM-like configurations superimposed on each other. This has also been explored directly from DFT calculations in other studies. For instance, in Ref.[\onlinecite{Dong2015}] DLM configurations combined with fluctuating moments and statistical models were used to calculate high temperature properties such as elastic constants and average magnetic moment in Fe. As a contrast, in the simulations employed here, all these fluctuations are naturally arising as outputs. The site resolved LSF energy shows very different behaviour for Ni and Fe atoms. As in DLM, the positive curvature of Ni means that it prefers not to sustain any magnetic moments on its own, nevertheless any finite Ni moments are stabilized by the entropy of spin fluctuations. On the other hand, the LSF energy of Fe has a minimum around 2.8 $\mu_B$ and as a consequence, the Fe moments are magnetically more resilient than Ni, and which in turn are induced from the Fe moments. It is worth noting that the individual output LSF energies on both Fe and Ni in Py also follow the Landau-type expansion as expected from the model.

\subsection{\label{subsec:Alm}Average magnetization, local moment and moment distribution}

\begin{figure}[htp]
\subfloat{\includegraphics[clip,width=0.85\columnwidth]{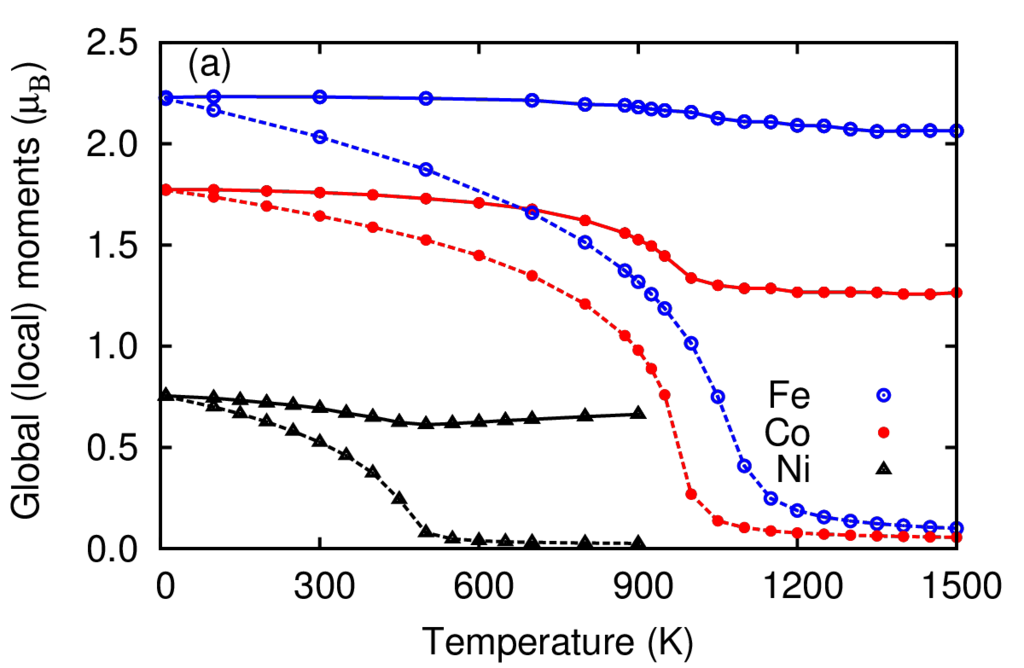}%
}

\subfloat{\includegraphics[clip,width=0.85\columnwidth]{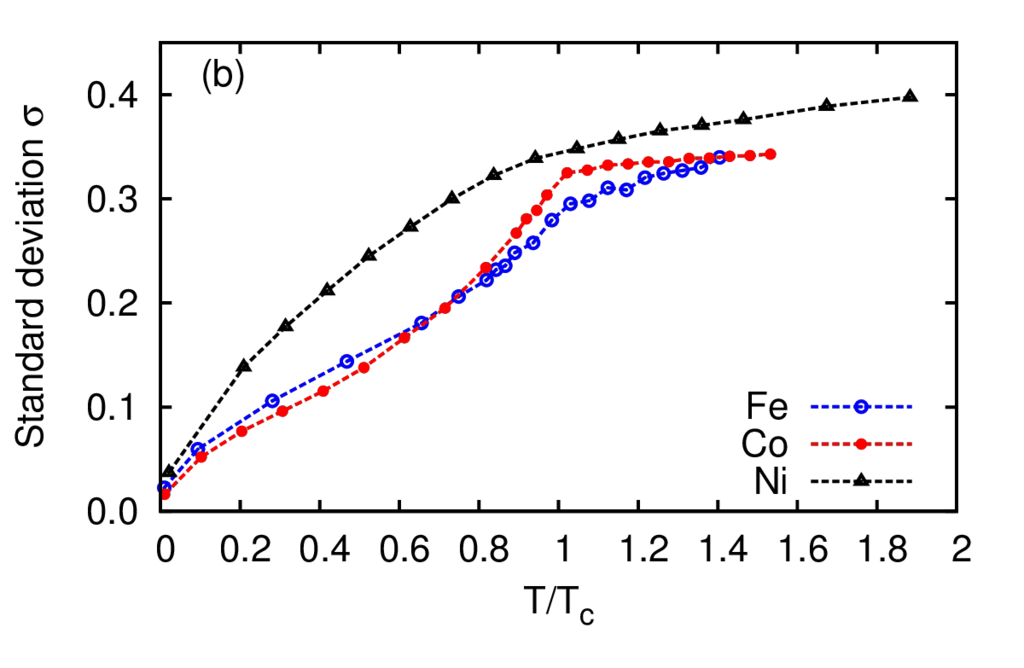}%
}
\caption{(a) The averaged global magnetic moment \textit{i.e} magnetization with dashed lines and the mean value of the local moments amplitude with solid lines. (b)  The standard deviation of the local moments  after normalized by the corresponding moment size at the zero temperature.}

\label{Mag_mmag_dev}
\end{figure}

The extended Heisenberg model where LSF is taken into account allows us to examine the dependency of the change in the local moment size on temperatures by performing the MC simulations. The magnetization and the local moments size with respect to temperature are displayed in Fig.~\ref{Mag_mmag_dev}(a) for Fe, Co and Ni.
We observe a linear reduction on the magnetization at low temperature in all the considered cases which does not follow the Bloch's $T^{3/2}$ as expected from using classical statistics.
However, here we are focusing at the high temperatures regime for temperature around $T_c$ and above where the classical spin model is well justified. The average moment size of Fe is reduced only weekly of about $10\%$ up to $T_c$. In contrast, the reduction of the average moments of Co and Ni in percentage increase by a factor of two compared to Fe. 
Our results are  consistent with what have been found in Ref.[\onlinecite{Rosengaard1997}] where the changes on the Ni size is continuous at \textit{Tc}, however a different behaviour for Ni was shown in Ref.[\onlinecite{Ruban2007}]. 
Despite of a rather constant average moment size, the variation on the level of individual local moment sizes is rather different. This is shown in the Fig.~\ref{Mag_mmag_dev}(b), where we have calculated the standard deviation of the local moment size as a function of temperature. Compare to Fe and Co where the standard deviation are relatively small at low temperatures, Ni has a large standard deviation even at low temperatures. This fact indicates that the description of a ``rigid moment'' in the longitudinal direction is appropriate for Fe, intermediate for Co but not so for Ni, as expected since Ni moments being the most itinerant of the three.


\begin{figure}[htb]
\includegraphics[width=.95\linewidth]{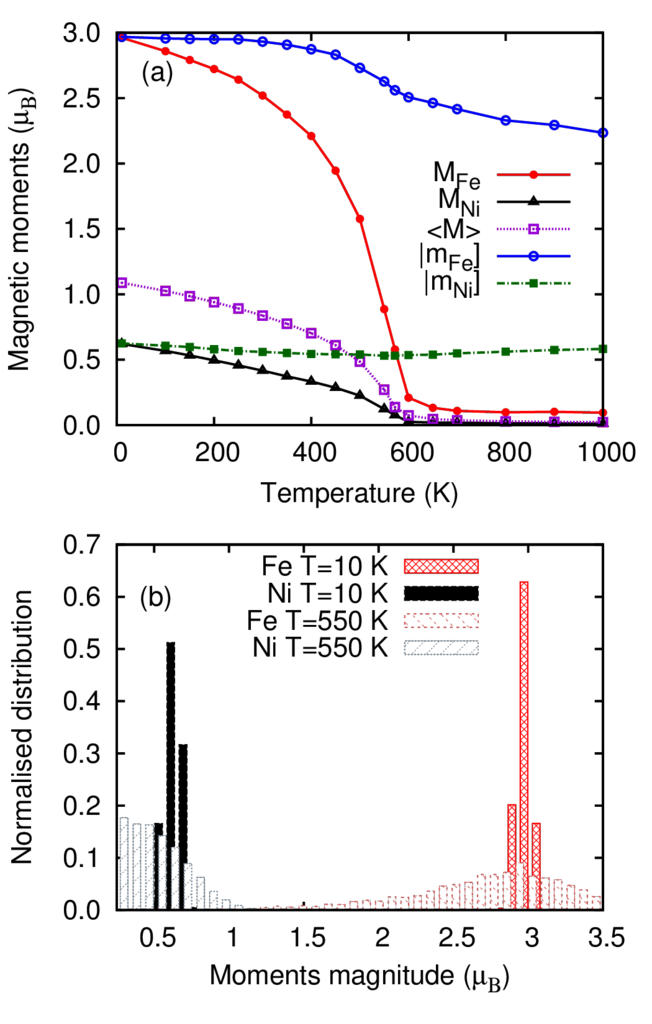}
\caption{(a) The average magnetization of Py $\langle M\rangle$, the sub-lattice magnetization of Fe and Ni ($M_{Fe}$ and $M_{Ni}$) and local moment size of Fe and Ni ($|m_{Fe}|$ and $|m_{Ni}|$) as a function of temperature. (b) Moment size distribution of Fe and Ni moments in Py at $T=10\ K$ and $T=550\ K$.}
\label{fig:Py_magnetization_hist}
\end{figure}

In Fig.~\ref{fig:Py_magnetization_hist}(a), the component-resolved local moment size and magnetization are calculated for Py. We focus on a few intricate features that could only be displayed in random alloys. First of all, the size of Fe moments in Py is calculated to be larger than it is in the elemental Fe while the moment size of Ni has roughly the same value, as expected from other studies\cite{Pan2016}.
Moreover, compared to the elemental Fe where moment sizes are rather constant, the local Fe moments in Py are more strongly reduced, in particular above $T_c$. Fig.~\ref{fig:Py_magnetization_hist}(b) the moment size distributions of Fe and Ni in Py are shown for two temperatures, $T=10\ K$ and $T=550\ K$. At $T=10\ K$, the longitudinal fluctuations are small for both Fe and Ni resulting in narrow distribution. At elevated temperatures, the longitudinal fluctuations are more energetically favored. At $T=550\ K$, right below the $T_c$ of the system, the moment distribution of Fe shows an asymmetry with a long tail to small moments. The distribution profile of Ni is rather different than Fe. It has a significant shift to small moments indicating that Ni prefer to eventually lose its moment.

\subsection{\label{subsec:MSRO}Impact of LSF on the high temperature properties}
\begin{figure}[htb]
\includegraphics[width=.95\linewidth]{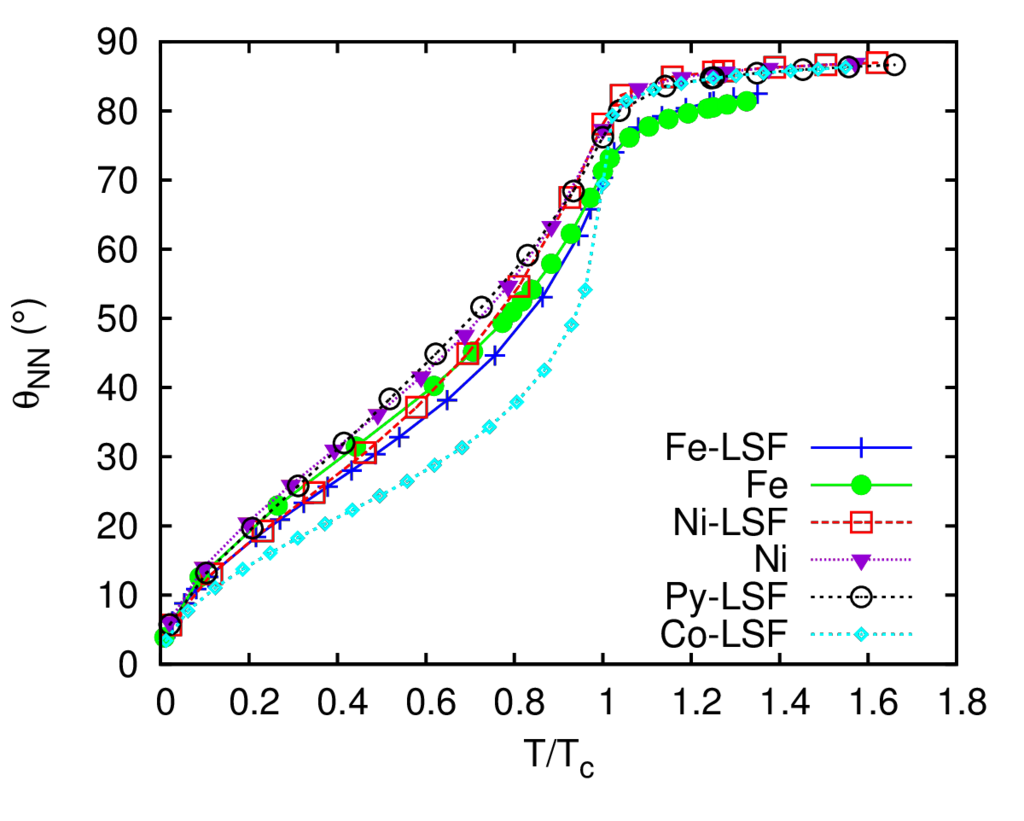}
\caption{The average tilting angle $\theta_{NN}$ between the nearest neighbours of Fe, Ni, Py and Co with selected cases included LSF term plot as a function of relative temperature $T/T_c$, where $T_c$ is the Curie temperature. LSF denotes that longitudinal fluctuations are included.}
\label{fig:NNangle_Fe_Ni_Py_Co}
\end{figure}

In Fig.~\ref{fig:NNangle_Fe_Ni_Py_Co} we show the magnetic short-range ordering (MSRO) in terms of the average tilting angle between the nearest neighbour moments as a function of the temperature relatively to the ordering temperature $T_c$. It is found that above $T_c$ the LSF term has negligible effect on the tilting angle, moreover it is only determined by the underlying structures, which can be seen from the fact that all systems in fcc lattices the temperature dependent $\theta_{NN}$ is essentially the same. Our results confirm the conclusion in Ref.~[\onlinecite{Wysocki2008}] where it was reported that the MSRO is expected to be weak for spin models of similar type as we employ in this study. In fact, the degree of MSRO is expected to be more or less constant for a large range of itinerancy. To be more specific, we obtain $cos\theta_{NN} = 0.2$ at $T =1.1T_c$ for Fe which is in an excellent agreement with the value calculated in Ref.~[\onlinecite{Wysocki2008}].
At temperatures below $T_c$, in the case of Fe and Ni, including LSF renders a reduced tiling angle compare to without. Co sticks out from the other systems by showing a rather different behavior where the $\theta_{NN}$ increases gradually at the low temperature regime while rapidly at around $T_c$.

\begin{figure}[htb]
\includegraphics[width=.95\linewidth]{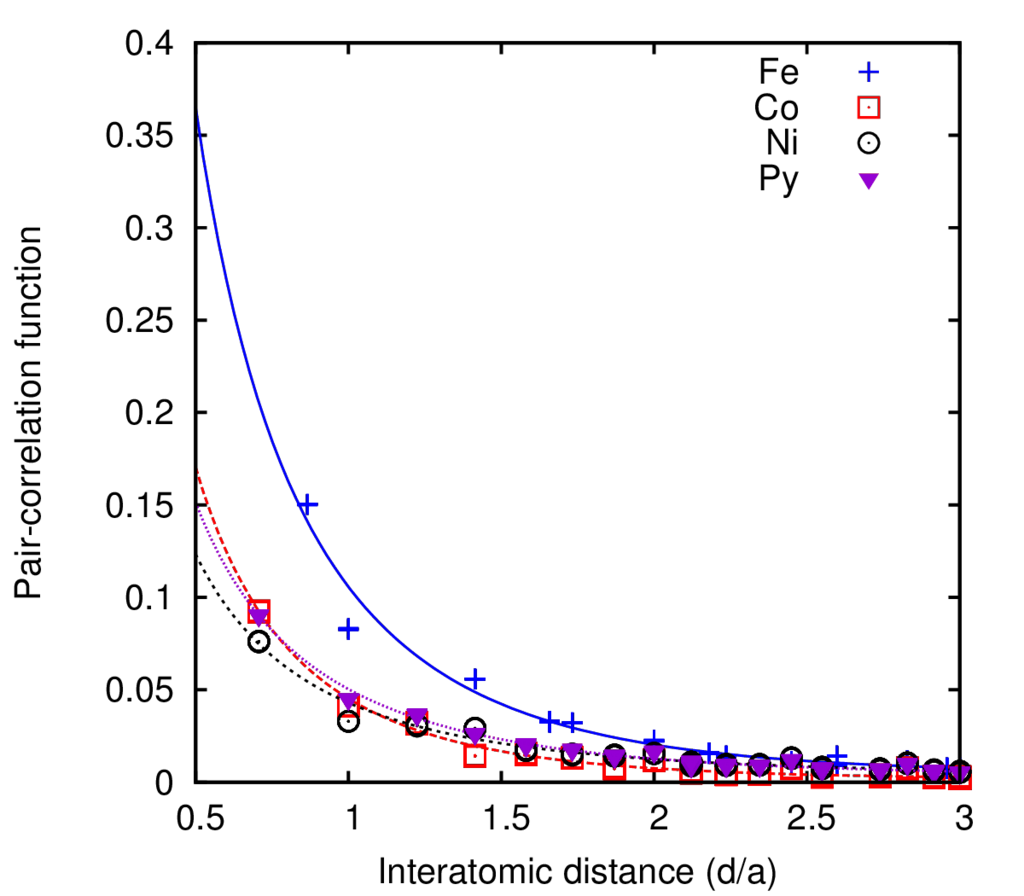}
\caption{Real space pair-correlation function $\langle \mathbf{m}_0 \cdot \mathbf{m}_j\rangle$, where $\mathbf{m}_0=\mathbf{M}_0/|M_0|$, shown as a function of interatomic distance $d$ relative to the lattice constant $a$ at $T=1.25T_c$. The dots are the calculated data points for Fe, Co, Ni and Py, respectively while the lines are plotted from the data fitted to the Ornstein-Zernike expression\cite{OZ1914}. Using this expression the inverse correlation length $\kappa$ can be extracted.}
\label{fig:Gr_DLM_wLSF}
\end{figure}

Apart from the average tilting angle that reveals only the nearest neighbour ordering, the magnetic order can also be evaluated through the correlation length of the real space correlation function $\langle \mathbf{m}_0 \cdot \mathbf{m}_j\rangle$ fitting to the Ornstein-Zernike expression\cite{OZ1914} $\sim exp(-\kappa d)/d$, where $d$ is the intersite distance and $\kappa$ (\AA$^{-1}$) the inverse correlation length. A comparison of the correlation length may not be elucidated directly from Fig.~\ref{fig:Gr_DLM_wLSF}, however $\kappa$ can be analyzed from the fitting of the Ornstein-Zernike expression.  
In Tab.~\ref{table:NNagle_Correlationlength} the summarized inverse correlation length $\kappa$ obtained for $T/T_c =1.25$  are listed. Ni is found having the longest correlation length in comparison with the other systems. Compared with the  previous theoretical work\cite{Rosengaard1997},  we obtained a good agreement in general. Despite we have taken into account both contributions from the nearest- and the next nearest-shells, whereas it did not in Ref.~[\onlinecite{Rosengaard1997}], the $\kappa$ is proved to be insensitive to this aspect. The surprisingly good agreement with the experimental value however, should not be over-credited. Due to small finite size effects present in the simulation data, we included a constant term in the least-square fitting procedure that shifts the entire function. In addition, while the monotonous decease of the Ornstein-Zernike expression is suitable for describing the long-range tail of the correlation function, it is unable to describe the oscillation noises due to the underlying crystal structure. The cumulative effects from those factors give rise to an estimated error bar of about $\pm 10 \%$.

\begin{table}[htb]
\begin{tabular}{@{\extracolsep{\fill}}c c c| c c c c}
\toprule
       &\multicolumn{2}{c|}{$\theta_{NN}$ ($^{\circ}$)} & \multicolumn{4}{c}{$\kappa$ (\AA$^{-1}$) at $T=1.25 T_c$ }\\
System & $T_c$ & 1.25 $T_c$ & LSF & N & Ref. & Exp.\\
\hline
Fe & 70 & 81 & 0.40 & 0.36 & 0.38$^a$ & 0.40 $^b$ \\
Co & 69 & 85 & 0.38 & 0.25 & 0.34$^a$ & -\\
Ni & 78 & 86 & 0.24 & 0.28 & 0.28$^a$ & 0.24 $^c$ \\
Py & 76 & 85 & 0.26 & 0.27 & - & - \\

\toprule
$^a$ Reference \onlinecite{Rosengaard1997}\\
$^b$ Reference \onlinecite{PhysRevB.33.1881}\\
$^c$ Reference \onlinecite{PhysRevB.30.2377}
\end{tabular}
\caption{The averaged tilted angle between the nearest neighbours $\theta_{NN}$ at $T_c$ and at $T=1.25T_c$ and correlation length $\kappa$ at $T=1.25T_c$. Our calculation including (without) LSF are denoted LSF (N) and compared to previous theory (Ref.) and experiments (Exp.).}
\label{table:NNagle_Correlationlength}
\end{table}

\subsection{Remarks of treatment of exchange, phase space measure and connection to previous works}\label{sec:remarks}

In this section, we will discuss in more details about some technical challenges and design choices in our implementation and comparison with previous works. From the outset, our implementation was designed to treat arbitrary number of magnetic components which give us some freedom for single-component systems such as Fe, Co and Ni. These systems could in our implementation be treated as binary alloys which could slightly affect the energetics of the exchange and LSF terms within each trial update in the Monte Carlo simulations. As mentioned in Sec.~\ref{sec:alsf}, the parameters in the simulations depend on the magnitude of the trial moment as well as the average magnitude of the surrounding moments of each of the other components. If there is only a single component, then it reduces to the local moment magnitude and the LSF term is nothing else than the Landau-type expression that was employed, for example, in Ref.[\onlinecite{Rosengaard1997}] for Fe,Co and Ni, [\onlinecite{Ma2012a}] for Fe and in [\onlinecite{Derlet2012}] for FeRh. Using the rescaled exchange parameters from LDA, we repeat calculations of $T_c$ for Fe and Ni but as treated as single component. We obtained 935~K and 308~K, respectively, compared to 926~K and 432~K as from Tab.~\ref{table:tc} treated as binary. Fe is basically unaffected by the simplified treatment due to the stable moments in which the LSF energy has a deep minimum, whereas Ni is found to be even softer than using the more advanced binary component treatment. If instead the ``bare" exchange interactions from the DLM state without rescaling are employed, the values of $T_c$ changes to 1120~K and 617~K for Fe and Ni, respectively. It is worth noting that the total energy in the FM limit is however not correct in this case. The value of Fe then becomes slightly overestimated compared to experiments (1043~K) while for Ni, excellent results that are remarkable close to the experimental value of 633~K is found. This rather good agreement can only be considered as fortuitous where shortcomings of the energetics from the underlying LDA (or GGA) calculations are balancing out each other. The value of Ni is also remarkable close to what was found in Ref.[\onlinecite{Ruban2007}] were similar exchange interactions from the DLM state was employed, but in that study a more elaborate LSF energy term was employed which indeed may improve the description of the high temperature properties. 

As pointed out in Section.~\ref{subsec:details}, the choice of PSM does matter in models including longitudinal spin fluctuations. Using the PSM including Jacobian weight instead of uniform put larger statistical weight to states with larger moments and in general that causes larger resistance to decreasing the moment. As a consequence, for all systems, apart from Fe that shows minor differences due to its stable moment, the critical temperature are considerably higher with the Jacobian included. For instance, the $T_c$ of Ni (LDA) is increased from 432~K to 613~K and in Py from 482~K to 708~K bringing the values much closer to experiment (633~K and 850~K, respectively). A more detailed comparison of different choices of PSM in the simulations is however left for a future study.

\section{Summary} \label{sec:summary}

We have constructed a general framework for atomistic simulations of multi-component random alloys including longitudinal fluctuations. The main ingredients consist of mapping total energies from electronic structure calculations for a number of fixed magnetic moments in both the low temperature ferromagnetic limit and the high temperature paramagnetic limit.  The model is then ``exact" in the both limits and interpolating between the two at intermediate temperatures. Within the model, each magnetic moment magnitude is allowed to change in a fashion that approximately corresponds to Stoner excitations in real materials and the model is constructed in such a  way that the strength of the exchange energy is independent on the starting reference state of the mapping, which is typically not the case found in previous studies. 

The computational framework has been implemented in Monte Carlo simulations and applied to the elemental transition metals Fe,Co and Ni, together with the binary random alloys Py and Fe-Co alloys. Regarding Curie temperatures, with the exception of Fe, the calculated values in general are slightly underestimated compared to experiments. However, the values are rather insensitive from the starting reference state which is an attractive feature lacking in ``normal" simulations for most materials. The simulations do qualitatively describe rather well the high temperature magnetic properties such as the correlation where excellent results compared to experiments are found.

\begin{acknowledgments}
The authors acknowledge valuable discussions with Levente Vitos. The work was financed through the VR (Swedish Research Council), SeRC (Swedish e-Science Research Centre) and GGS (G\"oran Gustafssons Stiftelser). A.B. acknowledges support from eSSENCE and the CEA-Enhanced Eurotalents programme, co-funded by FP7 Marie Skłodowska-Curie COFUND Programme (Grant Agreement n° 600382). The computations were performed on resources provided by SNIC (Swedish National Infrastructure for computing) at NSC (National Supercomputer Centre) in Link\"oping, Sweden.
\end{acknowledgments}

\bibliography{LSF_REF}

\begin{thebibliography}{59}%
\makeatletter
\providecommand \@ifxundefined [1]{%
 \@ifx{#1\undefined}
}%
\providecommand \@ifnum [1]{%
 \ifnum #1\expandafter \@firstoftwo
 \else \expandafter \@secondoftwo
 \fi
}%
\providecommand \@ifx [1]{%
 \ifx #1\expandafter \@firstoftwo
 \else \expandafter \@secondoftwo
 \fi
}%
\providecommand \natexlab [1]{#1}%
\providecommand \enquote  [1]{``#1''}%
\providecommand \bibnamefont  [1]{#1}%
\providecommand \bibfnamefont [1]{#1}%
\providecommand \citenamefont [1]{#1}%
\providecommand \href@noop [0]{\@secondoftwo}%
\providecommand \href [0]{\begingroup \@sanitize@url \@href}%
\providecommand \@href[1]{\@@startlink{#1}\@@href}%
\providecommand \@@href[1]{\endgroup#1\@@endlink}%
\providecommand \@sanitize@url [0]{\catcode `\\12\catcode `\$12\catcode
  `\&12\catcode `\#12\catcode `\^12\catcode `\_12\catcode `\%12\relax}%
\providecommand \@@startlink[1]{}%
\providecommand \@@endlink[0]{}%
\providecommand \url  [0]{\begingroup\@sanitize@url \@url }%
\providecommand \@url [1]{\endgroup\@href {#1}{\urlprefix }}%
\providecommand \urlprefix  [0]{URL }%
\providecommand \Eprint [0]{\href }%
\providecommand \doibase [0]{http://dx.doi.org/}%
\providecommand \selectlanguage [0]{\@gobble}%
\providecommand \bibinfo  [0]{\@secondoftwo}%
\providecommand \bibfield  [0]{\@secondoftwo}%
\providecommand \translation [1]{[#1]}%
\providecommand \BibitemOpen [0]{}%
\providecommand \bibitemStop [0]{}%
\providecommand \bibitemNoStop [0]{.\EOS\space}%
\providecommand \EOS [0]{\spacefactor3000\relax}%
\providecommand \BibitemShut  [1]{\csname bibitem#1\endcsname}%
\let\auto@bib@innerbib\@empty
\bibitem [{\citenamefont {Heisenberg}(1928)}]{Heisenberg1928}%
  \BibitemOpen
  \bibfield  {author} {\bibinfo {author} {\bibfnamefont {W.}~\bibnamefont
  {Heisenberg}},\ }\href {\doibase 10.1007/BF01328601} {\bibfield  {journal}
  {\bibinfo  {journal} {Zeitschrift f{\"u}r Physik}\ }\textbf {\bibinfo
  {volume} {49}},\ \bibinfo {pages} {619} (\bibinfo {year} {1928})}\BibitemShut
  {NoStop}%
\bibitem [{\citenamefont {Bloch}(1929)}]{Bloch1929}%
  \BibitemOpen
  \bibfield  {author} {\bibinfo {author} {\bibfnamefont {F.}~\bibnamefont
  {Bloch}},\ }\href {\doibase 10.1007/BF01340281} {\bibfield  {journal}
  {\bibinfo  {journal} {Zeitschrift f{\"u}r Physik}\ }\textbf {\bibinfo
  {volume} {57}},\ \bibinfo {pages} {545} (\bibinfo {year} {1929})}\BibitemShut
  {NoStop}%
\bibitem [{\citenamefont {Stoner}(1936)}]{Stoner1936}%
  \BibitemOpen
  \bibfield  {author} {\bibinfo {author} {\bibfnamefont {E.~C.}\ \bibnamefont
  {Stoner}},\ }\href {\doibase 10.1098/rspa.1936.0075} {\bibfield  {journal}
  {\bibinfo  {journal} {Proceedings of the Royal Society of London A:
  Mathematical, Physical and Engineering Sciences}\ }\textbf {\bibinfo {volume}
  {154}},\ \bibinfo {pages} {656} (\bibinfo {year} {1936})}\BibitemShut
  {NoStop}%
\bibitem [{\citenamefont {Slater}(1936{\natexlab{a}})}]{Slater1936}%
  \BibitemOpen
  \bibfield  {author} {\bibinfo {author} {\bibfnamefont {J.~C.}\ \bibnamefont
  {Slater}},\ }\href {\doibase 10.1103/PhysRev.49.537} {\bibfield  {journal}
  {\bibinfo  {journal} {Phys. Rev.}\ }\textbf {\bibinfo {volume} {49}},\
  \bibinfo {pages} {537} (\bibinfo {year} {1936}{\natexlab{a}})}\BibitemShut
  {NoStop}%
\bibitem [{\citenamefont {Slater}(1936{\natexlab{b}})}]{Slater1936b}%
  \BibitemOpen
  \bibfield  {author} {\bibinfo {author} {\bibfnamefont {J.~C.}\ \bibnamefont
  {Slater}},\ }\href {\doibase 10.1103/PhysRev.49.931} {\bibfield  {journal}
  {\bibinfo  {journal} {Phys. Rev.}\ }\textbf {\bibinfo {volume} {49}},\
  \bibinfo {pages} {931} (\bibinfo {year} {1936}{\natexlab{b}})}\BibitemShut
  {NoStop}%
\bibitem [{\citenamefont {Stoner}(1938)}]{Stoner1938}%
  \BibitemOpen
  \bibfield  {author} {\bibinfo {author} {\bibfnamefont {E.~C.}\ \bibnamefont
  {Stoner}},\ }\href {\doibase 10.1098/rspa.1938.0066} {\bibfield  {journal}
  {\bibinfo  {journal} {Proceedings of the Royal Society of London A:
  Mathematical, Physical and Engineering Sciences}\ }\textbf {\bibinfo {volume}
  {165}},\ \bibinfo {pages} {372} (\bibinfo {year} {1938})}\BibitemShut
  {NoStop}%
\bibitem [{\citenamefont {Moriya}(1979)}]{Moriya1979}%
  \BibitemOpen
  \bibfield  {author} {\bibinfo {author} {\bibfnamefont {T.}~\bibnamefont
  {Moriya}},\ }\href {\doibase http://dx.doi.org/10.1016/0304-8853(79)90201-4}
  {\bibfield  {journal} {\bibinfo  {journal} {Journal of Magnetism and Magnetic
  Materials}\ }\textbf {\bibinfo {volume} {14}},\ \bibinfo {pages} {1 }
  (\bibinfo {year} {1979})}\BibitemShut {NoStop}%
\bibitem [{\citenamefont {Uhl}\ and\ \citenamefont {K\"ubler}(1996)}]{Uhl1996}%
  \BibitemOpen
  \bibfield  {author} {\bibinfo {author} {\bibfnamefont {M.}~\bibnamefont
  {Uhl}}\ and\ \bibinfo {author} {\bibfnamefont {J.}~\bibnamefont {K\"ubler}},\
  }\href {\doibase 10.1103/PhysRevLett.77.334} {\bibfield  {journal} {\bibinfo
  {journal} {Phys. Rev. Lett.}\ }\textbf {\bibinfo {volume} {77}},\ \bibinfo
  {pages} {334} (\bibinfo {year} {1996})}\BibitemShut {NoStop}%
\bibitem [{\citenamefont {Rosengaard}\ and\ \citenamefont
  {Johansson}(1997)}]{Rosengaard1997}%
  \BibitemOpen
  \bibfield  {author} {\bibinfo {author} {\bibfnamefont {N.~M.}\ \bibnamefont
  {Rosengaard}}\ and\ \bibinfo {author} {\bibfnamefont {B.}~\bibnamefont
  {Johansson}},\ }\href {\doibase 10.1103/PhysRevB.55.14975} {\bibfield
  {journal} {\bibinfo  {journal} {Phys. Rev. B}\ }\textbf {\bibinfo {volume}
  {55}},\ \bibinfo {pages} {14975} (\bibinfo {year} {1997})}\BibitemShut
  {NoStop}%
\bibitem [{\citenamefont {Ruban}\ \emph {et~al.}(2007)\citenamefont {Ruban},
  \citenamefont {Khmelevskyi}, \citenamefont {Mohn},\ and\ \citenamefont
  {Johansson}}]{Ruban2007}%
  \BibitemOpen
  \bibfield  {author} {\bibinfo {author} {\bibfnamefont {A.~V.}\ \bibnamefont
  {Ruban}}, \bibinfo {author} {\bibfnamefont {S.}~\bibnamefont {Khmelevskyi}},
  \bibinfo {author} {\bibfnamefont {P.}~\bibnamefont {Mohn}}, \ and\ \bibinfo
  {author} {\bibfnamefont {B.}~\bibnamefont {Johansson}},\ }\href {\doibase
  10.1103/PhysRevB.75.054402} {\bibfield  {journal} {\bibinfo  {journal} {Phys.
  Rev. B}\ }\textbf {\bibinfo {volume} {75}},\ \bibinfo {pages} {054402}
  (\bibinfo {year} {2007})}\BibitemShut {NoStop}%
\bibitem [{\citenamefont {Mohn}\ \emph {et~al.}(1990)\citenamefont {Mohn},
  \citenamefont {Schwarz},\ and\ \citenamefont {Wagner}}]{Mohn1990}%
  \BibitemOpen
  \bibfield  {author} {\bibinfo {author} {\bibfnamefont {P.}~\bibnamefont
  {Mohn}}, \bibinfo {author} {\bibfnamefont {K.}~\bibnamefont {Schwarz}}, \
  and\ \bibinfo {author} {\bibfnamefont {D.}~\bibnamefont {Wagner}},\ }\href
  {\doibase http://dx.doi.org/10.1016/0921-4526(89)90122-1} {\bibfield
  {journal} {\bibinfo  {journal} {Physica B: Condensed Matter}\ }\textbf
  {\bibinfo {volume} {161}},\ \bibinfo {pages} {153 } (\bibinfo {year}
  {1990})}\BibitemShut {NoStop}%
\bibitem [{\citenamefont {Le\ifmmode \check{z}\else
  \v{z}\fi{}ai\ifmmode~\acute{c}\else \'{c}\fi{}}\ \emph
  {et~al.}(2006)\citenamefont {Le\ifmmode \check{z}\else
  \v{z}\fi{}ai\ifmmode~\acute{c}\else \'{c}\fi{}}, \citenamefont {Mavropoulos},
  \citenamefont {Enkovaara}, \citenamefont {Bihlmayer},\ and\ \citenamefont
  {Bl\"ugel}}]{Lezaic2006}%
  \BibitemOpen
  \bibfield  {author} {\bibinfo {author} {\bibfnamefont {M.}~\bibnamefont
  {Le\ifmmode \check{z}\else \v{z}\fi{}ai\ifmmode~\acute{c}\else \'{c}\fi{}}},
  \bibinfo {author} {\bibfnamefont {P.}~\bibnamefont {Mavropoulos}}, \bibinfo
  {author} {\bibfnamefont {J.}~\bibnamefont {Enkovaara}}, \bibinfo {author}
  {\bibfnamefont {G.}~\bibnamefont {Bihlmayer}}, \ and\ \bibinfo {author}
  {\bibfnamefont {S.}~\bibnamefont {Bl\"ugel}},\ }\href {\doibase
  10.1103/PhysRevLett.97.026404} {\bibfield  {journal} {\bibinfo  {journal}
  {Phys. Rev. Lett.}\ }\textbf {\bibinfo {volume} {97}},\ \bibinfo {pages}
  {026404} (\bibinfo {year} {2006})}\BibitemShut {NoStop}%
\bibitem [{\citenamefont {Le\ifmmode \check{z}\else
  \v{z}\fi{}ai\ifmmode~\acute{c}\else \'{c}\fi{}}\ \emph
  {et~al.}(2013)\citenamefont {Le\ifmmode \check{z}\else
  \v{z}\fi{}ai\ifmmode~\acute{c}\else \'{c}\fi{}}, \citenamefont {Mavropoulos},
  \citenamefont {Bihlmayer},\ and\ \citenamefont {Bl\"ugel}}]{Lezaic2013}%
  \BibitemOpen
  \bibfield  {author} {\bibinfo {author} {\bibfnamefont {M.}~\bibnamefont
  {Le\ifmmode \check{z}\else \v{z}\fi{}ai\ifmmode~\acute{c}\else \'{c}\fi{}}},
  \bibinfo {author} {\bibfnamefont {P.}~\bibnamefont {Mavropoulos}}, \bibinfo
  {author} {\bibfnamefont {G.}~\bibnamefont {Bihlmayer}}, \ and\ \bibinfo
  {author} {\bibfnamefont {S.}~\bibnamefont {Bl\"ugel}},\ }\href {\doibase
  10.1103/PhysRevB.88.134403} {\bibfield  {journal} {\bibinfo  {journal} {Phys.
  Rev. B}\ }\textbf {\bibinfo {volume} {88}},\ \bibinfo {pages} {134403}
  (\bibinfo {year} {2013})}\BibitemShut {NoStop}%
\bibitem [{\citenamefont {Sandratskii}\ \emph {et~al.}(2007)\citenamefont
  {Sandratskii}, \citenamefont {Singer},\ and\ \citenamefont {\ifmmode
  \mbox{\c{S}}\else \c{S}\fi{}a\ifmmode \mbox{\c{s}}\else \c{s}\fi{}\ifmmode
  \imath \else \i \fi{}o\ifmmode~\breve{g}\else
  \u{g}\fi{}lu}}]{Sandratskii2007}%
  \BibitemOpen
  \bibfield  {author} {\bibinfo {author} {\bibfnamefont {L.~M.}\ \bibnamefont
  {Sandratskii}}, \bibinfo {author} {\bibfnamefont {R.}~\bibnamefont {Singer}},
  \ and\ \bibinfo {author} {\bibfnamefont {E.}~\bibnamefont {\ifmmode
  \mbox{\c{S}}\else \c{S}\fi{}a\ifmmode \mbox{\c{s}}\else \c{s}\fi{}\ifmmode
  \imath \else \i \fi{}o\ifmmode~\breve{g}\else \u{g}\fi{}lu}},\ }\href
  {\doibase 10.1103/PhysRevB.76.184406} {\bibfield  {journal} {\bibinfo
  {journal} {Phys. Rev. B}\ }\textbf {\bibinfo {volume} {76}},\ \bibinfo
  {pages} {184406} (\bibinfo {year} {2007})}\BibitemShut {NoStop}%
\bibitem [{\citenamefont {Sandratskii}(2008)}]{Sandratskii2008}%
  \BibitemOpen
  \bibfield  {author} {\bibinfo {author} {\bibfnamefont {L.~M.}\ \bibnamefont
  {Sandratskii}},\ }\href {\doibase 10.1103/PhysRevB.78.094425} {\bibfield
  {journal} {\bibinfo  {journal} {Phys. Rev. B}\ }\textbf {\bibinfo {volume}
  {78}},\ \bibinfo {pages} {094425} (\bibinfo {year} {2008})}\BibitemShut
  {NoStop}%
\bibitem [{\citenamefont {Polesya}\ \emph {et~al.}(2010)\citenamefont
  {Polesya}, \citenamefont {Mankovsky}, \citenamefont {Sipr}, \citenamefont
  {Meindl}, \citenamefont {Strunk},\ and\ \citenamefont {Ebert}}]{Polesya2010}%
  \BibitemOpen
  \bibfield  {author} {\bibinfo {author} {\bibfnamefont {S.}~\bibnamefont
  {Polesya}}, \bibinfo {author} {\bibfnamefont {S.}~\bibnamefont {Mankovsky}},
  \bibinfo {author} {\bibfnamefont {O.}~\bibnamefont {Sipr}}, \bibinfo {author}
  {\bibfnamefont {W.}~\bibnamefont {Meindl}}, \bibinfo {author} {\bibfnamefont
  {C.}~\bibnamefont {Strunk}}, \ and\ \bibinfo {author} {\bibfnamefont
  {H.}~\bibnamefont {Ebert}},\ }\href {\doibase 10.1103/PhysRevB.82.214409}
  {\bibfield  {journal} {\bibinfo  {journal} {Phys. Rev. B}\ }\textbf {\bibinfo
  {volume} {82}},\ \bibinfo {pages} {214409} (\bibinfo {year}
  {2010})}\BibitemShut {NoStop}%
\bibitem [{\citenamefont {Polesya}\ \emph {et~al.}(2016)\citenamefont
  {Polesya}, \citenamefont {Mankovsky}, \citenamefont {K\"odderitzsch},
  \citenamefont {Min\'ar},\ and\ \citenamefont {Ebert}}]{Polesya2016}%
  \BibitemOpen
  \bibfield  {author} {\bibinfo {author} {\bibfnamefont {S.}~\bibnamefont
  {Polesya}}, \bibinfo {author} {\bibfnamefont {S.}~\bibnamefont {Mankovsky}},
  \bibinfo {author} {\bibfnamefont {D.}~\bibnamefont {K\"odderitzsch}},
  \bibinfo {author} {\bibfnamefont {J.}~\bibnamefont {Min\'ar}}, \ and\
  \bibinfo {author} {\bibfnamefont {H.}~\bibnamefont {Ebert}},\ }\href
  {\doibase 10.1103/PhysRevB.93.024423} {\bibfield  {journal} {\bibinfo
  {journal} {Phys. Rev. B}\ }\textbf {\bibinfo {volume} {93}},\ \bibinfo
  {pages} {024423} (\bibinfo {year} {2016})}\BibitemShut {NoStop}%
\bibitem [{\citenamefont {Khmelevskyi}(2016{\natexlab{a}})}]{Khmelevskyi2016}%
  \BibitemOpen
  \bibfield  {author} {\bibinfo {author} {\bibfnamefont {S.}~\bibnamefont
  {Khmelevskyi}},\ }\href {\doibase 10.1103/PhysRevB.94.024420} {\bibfield
  {journal} {\bibinfo  {journal} {Phys. Rev. B}\ }\textbf {\bibinfo {volume}
  {94}},\ \bibinfo {pages} {024420} (\bibinfo {year}
  {2016}{\natexlab{a}})}\BibitemShut {NoStop}%
\bibitem [{\citenamefont {Ma}\ and\ \citenamefont {Dudarev}(2012)}]{Ma2012a}%
  \BibitemOpen
  \bibfield  {author} {\bibinfo {author} {\bibfnamefont {P.-W.}\ \bibnamefont
  {Ma}}\ and\ \bibinfo {author} {\bibfnamefont {S.~L.}\ \bibnamefont
  {Dudarev}},\ }\href {\doibase 10.1103/PhysRevB.86.054416} {\bibfield
  {journal} {\bibinfo  {journal} {Phys. Rev. B}\ }\textbf {\bibinfo {volume}
  {86}},\ \bibinfo {pages} {054416} (\bibinfo {year} {2012})}\BibitemShut
  {NoStop}%
\bibitem [{\citenamefont {Ma}\ and\ \citenamefont {Dudarev}(2014)}]{Ma2014}%
  \BibitemOpen
  \bibfield  {author} {\bibinfo {author} {\bibfnamefont {P.-W.}\ \bibnamefont
  {Ma}}\ and\ \bibinfo {author} {\bibfnamefont {S.~L.}\ \bibnamefont
  {Dudarev}},\ }\href {\doibase 10.1103/PhysRevB.90.024425} {\bibfield
  {journal} {\bibinfo  {journal} {Phys. Rev. B}\ }\textbf {\bibinfo {volume}
  {90}},\ \bibinfo {pages} {024425} (\bibinfo {year} {2014})}\BibitemShut
  {NoStop}%
\bibitem [{\citenamefont {Buczek}\ \emph {et~al.}(2011)\citenamefont {Buczek},
  \citenamefont {Ernst},\ and\ \citenamefont {Sandratskii}}]{Buczek2011}%
  \BibitemOpen
  \bibfield  {author} {\bibinfo {author} {\bibfnamefont {P.}~\bibnamefont
  {Buczek}}, \bibinfo {author} {\bibfnamefont {A.}~\bibnamefont {Ernst}}, \
  and\ \bibinfo {author} {\bibfnamefont {L.~M.}\ \bibnamefont {Sandratskii}},\
  }\href {\doibase 10.1103/PhysRevB.84.174418} {\bibfield  {journal} {\bibinfo
  {journal} {Phys. Rev. B}\ }\textbf {\bibinfo {volume} {84}},\ \bibinfo
  {pages} {174418} (\bibinfo {year} {2011})}\BibitemShut {NoStop}%
\bibitem [{\citenamefont {Lounis}\ \emph {et~al.}(2015)\citenamefont {Lounis},
  \citenamefont {dos Santos~Dias},\ and\ \citenamefont
  {Schweflinghaus}}]{Lounis2015}%
  \BibitemOpen
  \bibfield  {author} {\bibinfo {author} {\bibfnamefont {S.}~\bibnamefont
  {Lounis}}, \bibinfo {author} {\bibfnamefont {M.}~\bibnamefont {dos
  Santos~Dias}}, \ and\ \bibinfo {author} {\bibfnamefont {B.}~\bibnamefont
  {Schweflinghaus}},\ }\href {\doibase 10.1103/PhysRevB.91.104420} {\bibfield
  {journal} {\bibinfo  {journal} {Phys. Rev. B}\ }\textbf {\bibinfo {volume}
  {91}},\ \bibinfo {pages} {104420} (\bibinfo {year} {2015})}\BibitemShut
  {NoStop}%
\bibitem [{\citenamefont {Costa}\ \emph {et~al.}(2004)\citenamefont {Costa},
  \citenamefont {Muniz},\ and\ \citenamefont {Mills}}]{Costa2004}%
  \BibitemOpen
  \bibfield  {author} {\bibinfo {author} {\bibfnamefont {A.~T.}\ \bibnamefont
  {Costa}}, \bibinfo {author} {\bibfnamefont {R.~B.}\ \bibnamefont {Muniz}}, \
  and\ \bibinfo {author} {\bibfnamefont {D.~L.}\ \bibnamefont {Mills}},\ }\href
  {\doibase 10.1103/PhysRevB.69.064413} {\bibfield  {journal} {\bibinfo
  {journal} {Phys. Rev. B}\ }\textbf {\bibinfo {volume} {69}},\ \bibinfo
  {pages} {064413} (\bibinfo {year} {2004})}\BibitemShut {NoStop}%
\bibitem [{\citenamefont {Heine}\ \emph {et~al.}(1990)\citenamefont {Heine},
  \citenamefont {Liechtenstein},\ and\ \citenamefont {Mryasov}}]{Heine1990}%
  \BibitemOpen
  \bibfield  {author} {\bibinfo {author} {\bibfnamefont {V.}~\bibnamefont
  {Heine}}, \bibinfo {author} {\bibfnamefont {A.~I.}\ \bibnamefont
  {Liechtenstein}}, \ and\ \bibinfo {author} {\bibfnamefont {O.~N.}\
  \bibnamefont {Mryasov}},\ }\href
  {http://stacks.iop.org/0295-5075/12/i=6/a=013} {\bibfield  {journal}
  {\bibinfo  {journal} {EPL (Europhysics Letters)}\ }\textbf {\bibinfo {volume}
  {12}},\ \bibinfo {pages} {545} (\bibinfo {year} {1990})}\BibitemShut
  {NoStop}%
\bibitem [{\citenamefont {Halilov}\ \emph {et~al.}(1998)\citenamefont
  {Halilov}, \citenamefont {Eschrig}, \citenamefont {Perlov},\ and\
  \citenamefont {Oppeneer}}]{Halilov1998}%
  \BibitemOpen
  \bibfield  {author} {\bibinfo {author} {\bibfnamefont {S.~V.}\ \bibnamefont
  {Halilov}}, \bibinfo {author} {\bibfnamefont {H.}~\bibnamefont {Eschrig}},
  \bibinfo {author} {\bibfnamefont {A.~Y.}\ \bibnamefont {Perlov}}, \ and\
  \bibinfo {author} {\bibfnamefont {P.~M.}\ \bibnamefont {Oppeneer}},\ }\href
  {\doibase 10.1103/PhysRevB.58.293} {\bibfield  {journal} {\bibinfo  {journal}
  {Phys. Rev. B}\ }\textbf {\bibinfo {volume} {58}},\ \bibinfo {pages} {293}
  (\bibinfo {year} {1998})}\BibitemShut {NoStop}%
\bibitem [{\citenamefont {Szilva}\ \emph {et~al.}(2013)\citenamefont {Szilva},
  \citenamefont {Costa}, \citenamefont {Bergman}, \citenamefont {Szunyogh},
  \citenamefont {Nordstr\"om},\ and\ \citenamefont {Eriksson}}]{Szilva2013}%
  \BibitemOpen
  \bibfield  {author} {\bibinfo {author} {\bibfnamefont {A.}~\bibnamefont
  {Szilva}}, \bibinfo {author} {\bibfnamefont {M.}~\bibnamefont {Costa}},
  \bibinfo {author} {\bibfnamefont {A.}~\bibnamefont {Bergman}}, \bibinfo
  {author} {\bibfnamefont {L.}~\bibnamefont {Szunyogh}}, \bibinfo {author}
  {\bibfnamefont {L.}~\bibnamefont {Nordstr\"om}}, \ and\ \bibinfo {author}
  {\bibfnamefont {O.}~\bibnamefont {Eriksson}},\ }\href {\doibase
  10.1103/PhysRevLett.111.127204} {\bibfield  {journal} {\bibinfo  {journal}
  {Phys. Rev. Lett.}\ }\textbf {\bibinfo {volume} {111}},\ \bibinfo {pages}
  {127204} (\bibinfo {year} {2013})}\BibitemShut {NoStop}%
\bibitem [{\citenamefont {Shallcross}\ \emph {et~al.}(2005)\citenamefont
  {Shallcross}, \citenamefont {Kissavos}, \citenamefont {Meded},\ and\
  \citenamefont {Ruban}}]{Shallcross2005}%
  \BibitemOpen
  \bibfield  {author} {\bibinfo {author} {\bibfnamefont {S.}~\bibnamefont
  {Shallcross}}, \bibinfo {author} {\bibfnamefont {A.~E.}\ \bibnamefont
  {Kissavos}}, \bibinfo {author} {\bibfnamefont {V.}~\bibnamefont {Meded}}, \
  and\ \bibinfo {author} {\bibfnamefont {A.~V.}\ \bibnamefont {Ruban}},\ }\href
  {\doibase 10.1103/PhysRevB.72.104437} {\bibfield  {journal} {\bibinfo
  {journal} {Phys. Rev. B}\ }\textbf {\bibinfo {volume} {72}},\ \bibinfo
  {pages} {104437} (\bibinfo {year} {2005})}\BibitemShut {NoStop}%
\bibitem [{\citenamefont {Drautz}\ and\ \citenamefont
  {F\"ahnle}(2004)}]{Drautz2004}%
  \BibitemOpen
  \bibfield  {author} {\bibinfo {author} {\bibfnamefont {R.}~\bibnamefont
  {Drautz}}\ and\ \bibinfo {author} {\bibfnamefont {M.}~\bibnamefont
  {F\"ahnle}},\ }\href {\doibase 10.1103/PhysRevB.69.104404} {\bibfield
  {journal} {\bibinfo  {journal} {Phys. Rev. B}\ }\textbf {\bibinfo {volume}
  {69}},\ \bibinfo {pages} {104404} (\bibinfo {year} {2004})}\BibitemShut
  {NoStop}%
\bibitem [{\citenamefont {Ruban}\ \emph {et~al.}(2002)\citenamefont {Ruban},
  \citenamefont {Simak}, \citenamefont {Korzhavyi},\ and\ \citenamefont
  {Skriver}}]{Ruban2002}%
  \BibitemOpen
  \bibfield  {author} {\bibinfo {author} {\bibfnamefont {A.~V.}\ \bibnamefont
  {Ruban}}, \bibinfo {author} {\bibfnamefont {S.~I.}\ \bibnamefont {Simak}},
  \bibinfo {author} {\bibfnamefont {P.~A.}\ \bibnamefont {Korzhavyi}}, \ and\
  \bibinfo {author} {\bibfnamefont {H.~L.}\ \bibnamefont {Skriver}},\ }\href
  {\doibase 10.1103/PhysRevB.66.024202} {\bibfield  {journal} {\bibinfo
  {journal} {Phys. Rev. B}\ }\textbf {\bibinfo {volume} {66}},\ \bibinfo
  {pages} {024202} (\bibinfo {year} {2002})}\BibitemShut {NoStop}%
\bibitem [{\citenamefont {Perdew}\ \emph {et~al.}(1996)\citenamefont {Perdew},
  \citenamefont {Burke},\ and\ \citenamefont {Ernzerhof}}]{Perdew1996}%
  \BibitemOpen
  \bibfield  {author} {\bibinfo {author} {\bibfnamefont {J.~P.}\ \bibnamefont
  {Perdew}}, \bibinfo {author} {\bibfnamefont {K.}~\bibnamefont {Burke}}, \
  and\ \bibinfo {author} {\bibfnamefont {M.}~\bibnamefont {Ernzerhof}},\ }\href
  {\doibase 10.1103/PhysRevLett.77.3865} {\bibfield  {journal} {\bibinfo
  {journal} {Phys. Rev. Lett.}\ }\textbf {\bibinfo {volume} {77}},\ \bibinfo
  {pages} {3865} (\bibinfo {year} {1996})}\BibitemShut {NoStop}%
\bibitem [{\citenamefont {Dederichs}\ \emph {et~al.}(1984)\citenamefont
  {Dederichs}, \citenamefont {Bl\"ugel}, \citenamefont {Zeller},\ and\
  \citenamefont {Akai}}]{Dederichs1984}%
  \BibitemOpen
  \bibfield  {author} {\bibinfo {author} {\bibfnamefont {P.~H.}\ \bibnamefont
  {Dederichs}}, \bibinfo {author} {\bibfnamefont {S.}~\bibnamefont {Bl\"ugel}},
  \bibinfo {author} {\bibfnamefont {R.}~\bibnamefont {Zeller}}, \ and\ \bibinfo
  {author} {\bibfnamefont {H.}~\bibnamefont {Akai}},\ }\href {\doibase
  10.1103/PhysRevLett.53.2512} {\bibfield  {journal} {\bibinfo  {journal}
  {Phys. Rev. Lett.}\ }\textbf {\bibinfo {volume} {53}},\ \bibinfo {pages}
  {2512} (\bibinfo {year} {1984})}\BibitemShut {NoStop}%
\bibitem [{\citenamefont {Schwarz}\ and\ \citenamefont
  {Mohn}(1984)}]{Schwarz1984}%
  \BibitemOpen
  \bibfield  {author} {\bibinfo {author} {\bibfnamefont {K.}~\bibnamefont
  {Schwarz}}\ and\ \bibinfo {author} {\bibfnamefont {P.}~\bibnamefont {Mohn}},\
  }\href {http://stacks.iop.org/0305-4608/14/i=7/a=008} {\bibfield  {journal}
  {\bibinfo  {journal} {Journal of Physics F: Metal Physics}\ }\textbf
  {\bibinfo {volume} {14}},\ \bibinfo {pages} {L129} (\bibinfo {year}
  {1984})}\BibitemShut {NoStop}%
\bibitem [{\citenamefont {Liechtenstein}\ \emph {et~al.}(1984)\citenamefont
  {Liechtenstein}, \citenamefont {Katsnelson},\ and\ \citenamefont
  {Gubanov}}]{LKAG1984}%
  \BibitemOpen
  \bibfield  {author} {\bibinfo {author} {\bibfnamefont {A.~I.}\ \bibnamefont
  {Liechtenstein}}, \bibinfo {author} {\bibfnamefont {M.~I.}\ \bibnamefont
  {Katsnelson}}, \ and\ \bibinfo {author} {\bibfnamefont {V.~A.}\ \bibnamefont
  {Gubanov}},\ }\href {http://stacks.iop.org/0305-4608/14/i=7/a=007} {\bibfield
   {journal} {\bibinfo  {journal} {Journal of Physics F: Metal Physics}\
  }\textbf {\bibinfo {volume} {14}},\ \bibinfo {pages} {L125} (\bibinfo {year}
  {1984})}\BibitemShut {NoStop}%
\bibitem [{\citenamefont {Liechtenstein}\ \emph {et~al.}(1987)\citenamefont
  {Liechtenstein}, \citenamefont {Katsnelson}, \citenamefont {Antropov},\ and\
  \citenamefont {Gubanov}}]{LKAG1987}%
  \BibitemOpen
  \bibfield  {author} {\bibinfo {author} {\bibfnamefont {A.}~\bibnamefont
  {Liechtenstein}}, \bibinfo {author} {\bibfnamefont {M.}~\bibnamefont
  {Katsnelson}}, \bibinfo {author} {\bibfnamefont {V.}~\bibnamefont
  {Antropov}}, \ and\ \bibinfo {author} {\bibfnamefont {V.}~\bibnamefont
  {Gubanov}},\ }\href {\doibase http://dx.doi.org/10.1016/0304-8853(87)90721-9}
  {\bibfield  {journal} {\bibinfo  {journal} {Journal of Magnetism and Magnetic
  Materials}\ }\textbf {\bibinfo {volume} {67}},\ \bibinfo {pages} {65 }
  (\bibinfo {year} {1987})}\BibitemShut {NoStop}%
\bibitem [{\citenamefont {Skubic}\ \emph {et~al.}(2008)\citenamefont {Skubic},
  \citenamefont {Hellsvik}, \citenamefont {Nordström},\ and\ \citenamefont
  {Eriksson}}]{Skubic2008}%
  \BibitemOpen
  \bibfield  {author} {\bibinfo {author} {\bibfnamefont {B.}~\bibnamefont
  {Skubic}}, \bibinfo {author} {\bibfnamefont {J.}~\bibnamefont {Hellsvik}},
  \bibinfo {author} {\bibfnamefont {L.}~\bibnamefont {Nordström}}, \ and\
  \bibinfo {author} {\bibfnamefont {O.}~\bibnamefont {Eriksson}},\ }\href
  {http://stacks.iop.org/0953-8984/20/i=31/a=315203} {\bibfield  {journal}
  {\bibinfo  {journal} {Journal of Physics: Condensed Matter}\ }\textbf
  {\bibinfo {volume} {20}},\ \bibinfo {pages} {315203} (\bibinfo {year}
  {2008})}\BibitemShut {NoStop}%
\bibitem [{\citenamefont {Eriksson}\ \emph {et~al.}(2017)\citenamefont
  {Eriksson}, \citenamefont {Bergman}, \citenamefont {Bergqvist},\ and\
  \citenamefont {Hellsvik}}]{ASDbook2017}%
  \BibitemOpen
  \bibfield  {author} {\bibinfo {author} {\bibfnamefont {O.}~\bibnamefont
  {Eriksson}}, \bibinfo {author} {\bibfnamefont {A.}~\bibnamefont {Bergman}},
  \bibinfo {author} {\bibfnamefont {L.}~\bibnamefont {Bergqvist}}, \ and\
  \bibinfo {author} {\bibfnamefont {J.}~\bibnamefont {Hellsvik}},\ }\href@noop
  {} {\emph {\bibinfo {title} {Atomistic Spin Dynamics: Foundations and
  Applications}}}\ (\bibinfo  {publisher} {Oxford University Press},\ \bibinfo
  {year} {2017})\BibitemShut {NoStop}%
\bibitem [{\citenamefont {Binder}(1981)}]{Binder1981}%
  \BibitemOpen
  \bibfield  {author} {\bibinfo {author} {\bibfnamefont {K.}~\bibnamefont
  {Binder}},\ }\href {\doibase 10.1007/BF01293604} {\bibfield  {journal}
  {\bibinfo  {journal} {Zeitschrift f{\"u}r Physik B Condensed Matter}\
  }\textbf {\bibinfo {volume} {43}},\ \bibinfo {pages} {119} (\bibinfo {year}
  {1981})}\BibitemShut {NoStop}%
\bibitem [{\citenamefont {Pajda}\ \emph {et~al.}(2001)\citenamefont {Pajda},
  \citenamefont {Kudrnovsk\'y}, \citenamefont {Turek}, \citenamefont {Drchal},\
  and\ \citenamefont {Bruno}}]{Pajda2001}%
  \BibitemOpen
  \bibfield  {author} {\bibinfo {author} {\bibfnamefont {M.}~\bibnamefont
  {Pajda}}, \bibinfo {author} {\bibfnamefont {J.}~\bibnamefont {Kudrnovsk\'y}},
  \bibinfo {author} {\bibfnamefont {I.}~\bibnamefont {Turek}}, \bibinfo
  {author} {\bibfnamefont {V.}~\bibnamefont {Drchal}}, \ and\ \bibinfo {author}
  {\bibfnamefont {P.}~\bibnamefont {Bruno}},\ }\href {\doibase
  10.1103/PhysRevB.64.174402} {\bibfield  {journal} {\bibinfo  {journal} {Phys.
  Rev. B}\ }\textbf {\bibinfo {volume} {64}},\ \bibinfo {pages} {174402}
  (\bibinfo {year} {2001})}\BibitemShut {NoStop}%
\bibitem [{\citenamefont {D\"urrenfeld}\ \emph {et~al.}(2015)\citenamefont
  {D\"urrenfeld}, \citenamefont {Gerhard}, \citenamefont {Chico}, \citenamefont
  {Dumas}, \citenamefont {Ranjbar}, \citenamefont {Bergman}, \citenamefont
  {Bergqvist}, \citenamefont {Delin}, \citenamefont {Gould}, \citenamefont
  {Molenkamp},\ and\ \citenamefont {\AA{}kerman}}]{Durrenfeld2015}%
  \BibitemOpen
  \bibfield  {author} {\bibinfo {author} {\bibfnamefont {P.}~\bibnamefont
  {D\"urrenfeld}}, \bibinfo {author} {\bibfnamefont {F.}~\bibnamefont
  {Gerhard}}, \bibinfo {author} {\bibfnamefont {J.}~\bibnamefont {Chico}},
  \bibinfo {author} {\bibfnamefont {R.~K.}\ \bibnamefont {Dumas}}, \bibinfo
  {author} {\bibfnamefont {M.}~\bibnamefont {Ranjbar}}, \bibinfo {author}
  {\bibfnamefont {A.}~\bibnamefont {Bergman}}, \bibinfo {author} {\bibfnamefont
  {L.}~\bibnamefont {Bergqvist}}, \bibinfo {author} {\bibfnamefont
  {A.}~\bibnamefont {Delin}}, \bibinfo {author} {\bibfnamefont
  {C.}~\bibnamefont {Gould}}, \bibinfo {author} {\bibfnamefont {L.~W.}\
  \bibnamefont {Molenkamp}}, \ and\ \bibinfo {author} {\bibfnamefont
  {J.}~\bibnamefont {\AA{}kerman}},\ }\href {\doibase
  10.1103/PhysRevB.92.214424} {\bibfield  {journal} {\bibinfo  {journal} {Phys.
  Rev. B}\ }\textbf {\bibinfo {volume} {92}},\ \bibinfo {pages} {214424}
  (\bibinfo {year} {2015})}\BibitemShut {NoStop}%
\bibitem [{\citenamefont {Murata}\ and\ \citenamefont
  {Doniach}(1972)}]{Murata1972}%
  \BibitemOpen
  \bibfield  {author} {\bibinfo {author} {\bibfnamefont {K.~K.}\ \bibnamefont
  {Murata}}\ and\ \bibinfo {author} {\bibfnamefont {S.}~\bibnamefont
  {Doniach}},\ }\href {\doibase 10.1103/PhysRevLett.29.285} {\bibfield
  {journal} {\bibinfo  {journal} {Phys. Rev. Lett.}\ }\textbf {\bibinfo
  {volume} {29}},\ \bibinfo {pages} {285} (\bibinfo {year} {1972})}\BibitemShut
  {NoStop}%
\bibitem [{\citenamefont {Wysocki}\ \emph {et~al.}(2008)\citenamefont
  {Wysocki}, \citenamefont {Glasbrenner},\ and\ \citenamefont
  {Belashchenko}}]{Wysocki2008}%
  \BibitemOpen
  \bibfield  {author} {\bibinfo {author} {\bibfnamefont {A.~L.}\ \bibnamefont
  {Wysocki}}, \bibinfo {author} {\bibfnamefont {J.~K.}\ \bibnamefont
  {Glasbrenner}}, \ and\ \bibinfo {author} {\bibfnamefont {K.~D.}\ \bibnamefont
  {Belashchenko}},\ }\href {\doibase 10.1103/PhysRevB.78.184419} {\bibfield
  {journal} {\bibinfo  {journal} {Phys. Rev. B}\ }\textbf {\bibinfo {volume}
  {78}},\ \bibinfo {pages} {184419} (\bibinfo {year} {2008})}\BibitemShut
  {NoStop}%
\bibitem [{\citenamefont {Dietermann}\ \emph {et~al.}(2012)\citenamefont
  {Dietermann}, \citenamefont {Sandratskii},\ and\ \citenamefont
  {Fähnle}}]{Dietermann2012}%
  \BibitemOpen
  \bibfield  {author} {\bibinfo {author} {\bibfnamefont {F.}~\bibnamefont
  {Dietermann}}, \bibinfo {author} {\bibfnamefont {L.}~\bibnamefont
  {Sandratskii}}, \ and\ \bibinfo {author} {\bibfnamefont {M.}~\bibnamefont
  {Fähnle}},\ }\href {\doibase http://dx.doi.org/10.1016/j.jmmm.2012.04.041}
  {\bibfield  {journal} {\bibinfo  {journal} {Journal of Magnetism and Magnetic
  Materials}\ }\textbf {\bibinfo {volume} {324}},\ \bibinfo {pages} {2693 }
  (\bibinfo {year} {2012})}\BibitemShut {NoStop}%
\bibitem [{\citenamefont {Ruban}\ \emph {et~al.}(2013)\citenamefont {Ruban},
  \citenamefont {Belonoshko},\ and\ \citenamefont {Skorodumova}}]{Ruban2013}%
  \BibitemOpen
  \bibfield  {author} {\bibinfo {author} {\bibfnamefont {A.~V.}\ \bibnamefont
  {Ruban}}, \bibinfo {author} {\bibfnamefont {A.~B.}\ \bibnamefont
  {Belonoshko}}, \ and\ \bibinfo {author} {\bibfnamefont {N.~V.}\ \bibnamefont
  {Skorodumova}},\ }\href {\doibase 10.1103/PhysRevB.87.014405} {\bibfield
  {journal} {\bibinfo  {journal} {Phys. Rev. B}\ }\textbf {\bibinfo {volume}
  {87}},\ \bibinfo {pages} {014405} (\bibinfo {year} {2013})}\BibitemShut
  {NoStop}%
\bibitem [{\citenamefont {Khmelevskyi}(2016{\natexlab{b}})}]{Sergii2016}%
  \BibitemOpen
  \bibfield  {author} {\bibinfo {author} {\bibfnamefont {S.}~\bibnamefont
  {Khmelevskyi}},\ }\href {\doibase 10.1103/PhysRevB.94.024420} {\bibfield
  {journal} {\bibinfo  {journal} {Phys. Rev. B}\ }\textbf {\bibinfo {volume}
  {94}},\ \bibinfo {pages} {024420} (\bibinfo {year}
  {2016}{\natexlab{b}})}\BibitemShut {NoStop}%
\bibitem [{\citenamefont {Wohlfarth}(1980)}]{Wohlfarth1980}%
  \BibitemOpen
  \bibfield  {author} {\bibinfo {author} {\bibfnamefont {E.~P.}\ \bibnamefont
  {Wohlfarth}},\ }\href@noop {} {\emph {\bibinfo {title} {Ferromagnetic
  Materials}}},\ edited by\ \bibinfo {editor} {\bibfnamefont {E.~P.}\
  \bibnamefont {Wohlfarth}}\ (\bibinfo  {publisher} {Nort-Holland, Amsterdam},\
  \bibinfo {year} {1980})\ p.~\bibinfo {pages} {1}\BibitemShut {NoStop}%
\bibitem [{\citenamefont {Le\ifmmode \check{z}\else
  \v{z}\fi{}ai\ifmmode~\acute{c}\else \'{c}\fi{}}\ \emph
  {et~al.}(2007)\citenamefont {Le\ifmmode \check{z}\else
  \v{z}\fi{}ai\ifmmode~\acute{c}\else \'{c}\fi{}}, \citenamefont {Mavropoulos},
  ,\ and\ \citenamefont {Bl\"ugel}}]{Lezaic2007}%
  \BibitemOpen
  \bibfield  {author} {\bibinfo {author} {\bibfnamefont {M.}~\bibnamefont
  {Le\ifmmode \check{z}\else \v{z}\fi{}ai\ifmmode~\acute{c}\else \'{c}\fi{}}},
  \bibinfo {author} {\bibfnamefont {P.}~\bibnamefont {Mavropoulos}}, , \ and\
  \bibinfo {author} {\bibfnamefont {S.}~\bibnamefont {Bl\"ugel}},\ }\href
  {\doibase 10.1063/1.2710181} {\bibfield  {journal} {\bibinfo  {journal}
  {Applied Physics Letters}\ }\textbf {\bibinfo {volume} {90}},\ \bibinfo
  {pages} {082504} (\bibinfo {year} {2007})}\BibitemShut {NoStop}%
\bibitem [{\citenamefont {Adachi}\ \emph {et~al.}(1991)\citenamefont {Adachi},
  \citenamefont {Bonnenberg}, \citenamefont {Franse}, \citenamefont {Gersdorf},
  \citenamefont {Hempel}, \citenamefont {Kanematsu}, \citenamefont {Misawa},
  \citenamefont {Shiga}, \citenamefont {Stearns},\ and\ \citenamefont
  {Wijn}}]{Landolt1991}%
  \BibitemOpen
  \bibfield  {author} {\bibinfo {author} {\bibfnamefont {K.}~\bibnamefont
  {Adachi}}, \bibinfo {author} {\bibfnamefont {D.}~\bibnamefont {Bonnenberg}},
  \bibinfo {author} {\bibfnamefont {J.~J.~M.}\ \bibnamefont {Franse}}, \bibinfo
  {author} {\bibfnamefont {R.}~\bibnamefont {Gersdorf}}, \bibinfo {author}
  {\bibfnamefont {K.~A.}\ \bibnamefont {Hempel}}, \bibinfo {author}
  {\bibfnamefont {K.}~\bibnamefont {Kanematsu}}, \bibinfo {author}
  {\bibfnamefont {S.}~\bibnamefont {Misawa}}, \bibinfo {author} {\bibfnamefont
  {M.}~\bibnamefont {Shiga}}, \bibinfo {author} {\bibfnamefont {M.~B.}\
  \bibnamefont {Stearns}}, \ and\ \bibinfo {author} {\bibfnamefont {H.~P.~J.}\
  \bibnamefont {Wijn}},\ }\href@noop {} {\emph {\bibinfo {title} {Magnetic
  Properties of Metals The Landold-B\"ornstein Database}}},\ edited by\
  \bibinfo {editor} {\bibfnamefont {H.~P.~J.}\ \bibnamefont {Wijn}},\ Vol.\
  \bibinfo {volume} {(New Series, Group III vol 19, Subvolume a)}\ (\bibinfo
  {publisher} {Springer, Berlin},\ \bibinfo {year} {1991})\BibitemShut
  {NoStop}%
\bibitem [{\citenamefont {Kudrnovsk\'y}\ \emph {et~al.}(2008)\citenamefont
  {Kudrnovsk\'y}, \citenamefont {Drchal},\ and\ \citenamefont
  {Bruno}}]{Kudrnovsky2008}%
  \BibitemOpen
  \bibfield  {author} {\bibinfo {author} {\bibfnamefont {J.}~\bibnamefont
  {Kudrnovsk\'y}}, \bibinfo {author} {\bibfnamefont {V.}~\bibnamefont
  {Drchal}}, \ and\ \bibinfo {author} {\bibfnamefont {P.}~\bibnamefont
  {Bruno}},\ }\href {\doibase 10.1103/PhysRevB.77.224422} {\bibfield  {journal}
  {\bibinfo  {journal} {Phys. Rev. B}\ }\textbf {\bibinfo {volume} {77}},\
  \bibinfo {pages} {224422} (\bibinfo {year} {2008})}\BibitemShut {NoStop}%
\bibitem [{\citenamefont {Kawahara}\ \emph {et~al.}(2003)\citenamefont
  {Kawahara}, \citenamefont {Iemura}, \citenamefont {Tsurekawa},\ and\
  \citenamefont {Watanabe}}]{Kawahara2003}%
  \BibitemOpen
  \bibfield  {author} {\bibinfo {author} {\bibfnamefont {K.}~\bibnamefont
  {Kawahara}}, \bibinfo {author} {\bibfnamefont {D.}~\bibnamefont {Iemura}},
  \bibinfo {author} {\bibfnamefont {S.}~\bibnamefont {Tsurekawa}}, \ and\
  \bibinfo {author} {\bibfnamefont {T.}~\bibnamefont {Watanabe}},\ }\href@noop
  {} {\bibfield  {journal} {\bibinfo  {journal} {Mater. Trans.}\ }\textbf
  {\bibinfo {volume} {44}},\ \bibinfo {pages} {2570} (\bibinfo {year}
  {2003})}\BibitemShut {NoStop}%
\bibitem [{\citenamefont {Abrikosov}\ \emph {et~al.}(1996)\citenamefont
  {Abrikosov}, \citenamefont {James}, \citenamefont {Eriksson}, \citenamefont
  {S\"oderlind}, \citenamefont {Ruban}, \citenamefont {Skriver},\ and\
  \citenamefont {Johansson}}]{Abrikosov1996}%
  \BibitemOpen
  \bibfield  {author} {\bibinfo {author} {\bibfnamefont {I.~A.}\ \bibnamefont
  {Abrikosov}}, \bibinfo {author} {\bibfnamefont {P.}~\bibnamefont {James}},
  \bibinfo {author} {\bibfnamefont {O.}~\bibnamefont {Eriksson}}, \bibinfo
  {author} {\bibfnamefont {P.}~\bibnamefont {S\"oderlind}}, \bibinfo {author}
  {\bibfnamefont {A.~V.}\ \bibnamefont {Ruban}}, \bibinfo {author}
  {\bibfnamefont {H.~L.}\ \bibnamefont {Skriver}}, \ and\ \bibinfo {author}
  {\bibfnamefont {B.}~\bibnamefont {Johansson}},\ }\href {\doibase
  10.1103/PhysRevB.54.3380} {\bibfield  {journal} {\bibinfo  {journal} {Phys.
  Rev. B}\ }\textbf {\bibinfo {volume} {54}},\ \bibinfo {pages} {3380}
  (\bibinfo {year} {1996})}\BibitemShut {NoStop}%
\bibitem [{\citenamefont {Stringfellow}(1968)}]{Stingfellow1968}%
  \BibitemOpen
  \bibfield  {author} {\bibinfo {author} {\bibfnamefont {M.~W.}\ \bibnamefont
  {Stringfellow}},\ }\href {http://stacks.iop.org/0022-3719/1/i=4/a=315}
  {\bibfield  {journal} {\bibinfo  {journal} {Journal of Physics C: Solid State
  Physics}\ }\textbf {\bibinfo {volume} {1}},\ \bibinfo {pages} {950} (\bibinfo
  {year} {1968})}\BibitemShut {NoStop}%
\bibitem [{\citenamefont {Mitchell}\ and\ \citenamefont
  {Paul}(1985)}]{Mitchell1985}%
  \BibitemOpen
  \bibfield  {author} {\bibinfo {author} {\bibfnamefont {P.~W.}\ \bibnamefont
  {Mitchell}}\ and\ \bibinfo {author} {\bibfnamefont {D.~M.}\ \bibnamefont
  {Paul}},\ }\href {\doibase 10.1103/PhysRevB.32.3272} {\bibfield  {journal}
  {\bibinfo  {journal} {Phys. Rev. B}\ }\textbf {\bibinfo {volume} {32}},\
  \bibinfo {pages} {3272} (\bibinfo {year} {1985})}\BibitemShut {NoStop}%
\bibitem [{\citenamefont {Liu}\ \emph {et~al.}(1994)\citenamefont {Liu},
  \citenamefont {Sooryakumar}, \citenamefont {Gutierrez},\ and\ \citenamefont
  {Prinz}}]{Liu1994}%
  \BibitemOpen
  \bibfield  {author} {\bibinfo {author} {\bibfnamefont {X.}~\bibnamefont
  {Liu}}, \bibinfo {author} {\bibfnamefont {R.}~\bibnamefont {Sooryakumar}},
  \bibinfo {author} {\bibfnamefont {C.~J.}\ \bibnamefont {Gutierrez}}, \ and\
  \bibinfo {author} {\bibfnamefont {G.~A.}\ \bibnamefont {Prinz}},\ }\href
  {\doibase http://dx.doi.org/10.1063/1.356763} {\bibfield  {journal} {\bibinfo
   {journal} {Journal of Applied Physics}\ }\textbf {\bibinfo {volume} {75}},\
  \bibinfo {pages} {7021} (\bibinfo {year} {1994})}\BibitemShut {NoStop}%
\bibitem [{\citenamefont {Dong}\ \emph {et~al.}(2015)\citenamefont {Dong},
  \citenamefont {Li}, \citenamefont {Sch\"onecker}, \citenamefont {Lu},
  \citenamefont {Chen},\ and\ \citenamefont {Vitos}}]{Dong2015}%
  \BibitemOpen
  \bibfield  {author} {\bibinfo {author} {\bibfnamefont {Z.}~\bibnamefont
  {Dong}}, \bibinfo {author} {\bibfnamefont {W.}~\bibnamefont {Li}}, \bibinfo
  {author} {\bibfnamefont {S.}~\bibnamefont {Sch\"onecker}}, \bibinfo {author}
  {\bibfnamefont {S.}~\bibnamefont {Lu}}, \bibinfo {author} {\bibfnamefont
  {D.}~\bibnamefont {Chen}}, \ and\ \bibinfo {author} {\bibfnamefont
  {L.}~\bibnamefont {Vitos}},\ }\href {\doibase 10.1103/PhysRevB.92.224420}
  {\bibfield  {journal} {\bibinfo  {journal} {Phys. Rev. B}\ }\textbf {\bibinfo
  {volume} {92}},\ \bibinfo {pages} {224420} (\bibinfo {year}
  {2015})}\BibitemShut {NoStop}%
\bibitem [{\citenamefont {Pan}\ \emph {et~al.}(2016)\citenamefont {Pan},
  \citenamefont {Chico}, \citenamefont {Hellsvik}, \citenamefont {Delin},
  \citenamefont {Bergman},\ and\ \citenamefont {Bergqvist}}]{Pan2016}%
  \BibitemOpen
  \bibfield  {author} {\bibinfo {author} {\bibfnamefont {F.}~\bibnamefont
  {Pan}}, \bibinfo {author} {\bibfnamefont {J.}~\bibnamefont {Chico}}, \bibinfo
  {author} {\bibfnamefont {J.}~\bibnamefont {Hellsvik}}, \bibinfo {author}
  {\bibfnamefont {A.}~\bibnamefont {Delin}}, \bibinfo {author} {\bibfnamefont
  {A.}~\bibnamefont {Bergman}}, \ and\ \bibinfo {author} {\bibfnamefont
  {L.}~\bibnamefont {Bergqvist}},\ }\href {\doibase 10.1103/PhysRevB.94.214410}
  {\bibfield  {journal} {\bibinfo  {journal} {Phys. Rev. B}\ }\textbf {\bibinfo
  {volume} {94}},\ \bibinfo {pages} {214410} (\bibinfo {year}
  {2016})}\BibitemShut {NoStop}%
\bibitem [{\citenamefont {Ornstein}\ and\ \citenamefont
  {Zernike}(1914)}]{OZ1914}%
  \BibitemOpen
  \bibfield  {author} {\bibinfo {author} {\bibfnamefont {L.}~\bibnamefont
  {Ornstein}}\ and\ \bibinfo {author} {\bibfnamefont {F.}~\bibnamefont
  {Zernike}},\ }\href@noop {} {\bibfield  {journal} {\bibinfo  {journal} {Proc.
  Acad. Sci. Amsterdam}\ }\textbf {\bibinfo {volume} {17}},\ \bibinfo {pages}
  {793} (\bibinfo {year} {1914})}\BibitemShut {NoStop}%
\bibitem [{\citenamefont {Shirane}\ \emph {et~al.}(1986)\citenamefont
  {Shirane}, \citenamefont {B\"oni},\ and\ \citenamefont
  {Wicksted}}]{PhysRevB.33.1881}%
  \BibitemOpen
  \bibfield  {author} {\bibinfo {author} {\bibfnamefont {G.}~\bibnamefont
  {Shirane}}, \bibinfo {author} {\bibfnamefont {P.}~\bibnamefont {B\"oni}}, \
  and\ \bibinfo {author} {\bibfnamefont {J.~P.}\ \bibnamefont {Wicksted}},\
  }\href {\doibase 10.1103/PhysRevB.33.1881} {\bibfield  {journal} {\bibinfo
  {journal} {Phys. Rev. B}\ }\textbf {\bibinfo {volume} {33}},\ \bibinfo
  {pages} {1881} (\bibinfo {year} {1986})}\BibitemShut {NoStop}%
\bibitem [{\citenamefont {Steinsvoll}\ \emph {et~al.}(1984)\citenamefont
  {Steinsvoll}, \citenamefont {Majkrzak}, \citenamefont {Shirane},\ and\
  \citenamefont {Wicksted}}]{PhysRevB.30.2377}%
  \BibitemOpen
  \bibfield  {author} {\bibinfo {author} {\bibfnamefont {O.}~\bibnamefont
  {Steinsvoll}}, \bibinfo {author} {\bibfnamefont {C.~F.}\ \bibnamefont
  {Majkrzak}}, \bibinfo {author} {\bibfnamefont {G.}~\bibnamefont {Shirane}}, \
  and\ \bibinfo {author} {\bibfnamefont {J.}~\bibnamefont {Wicksted}},\ }\href
  {\doibase 10.1103/PhysRevB.30.2377} {\bibfield  {journal} {\bibinfo
  {journal} {Phys. Rev. B}\ }\textbf {\bibinfo {volume} {30}},\ \bibinfo
  {pages} {2377} (\bibinfo {year} {1984})}\BibitemShut {NoStop}%
\bibitem [{\citenamefont {Derlet}(2012)}]{Derlet2012}%
  \BibitemOpen
  \bibfield  {author} {\bibinfo {author} {\bibfnamefont {P.~M.}\ \bibnamefont
  {Derlet}},\ }\href {\doibase 10.1103/PhysRevB.85.174431} {\bibfield
  {journal} {\bibinfo  {journal} {Phys. Rev. B}\ }\textbf {\bibinfo {volume}
  {85}},\ \bibinfo {pages} {174431} (\bibinfo {year} {2012})}\BibitemShut
  {NoStop}%
\end{thebibliography}%
\end{document}